# A Production-Ready Machine Learning System for Inclusive Employment: Requirements Engineering and Implementation of AI-Driven Disability Job Matching Platform

Oleksandr Kuznetsov [1,2*]

[1] Department of Theoretical and Applied Sciences, eCampus University, Via Isimbardi 10, Novedrate (CO), 22060, Italy

[2] Department of Intelligent Software Systems and Technologies, School of Computer Science and Artificial Intelligence, V.N. Karazin Kharkiv National University, 4 Svobody Sq., 61022 Kharkiv, Ukraine

Email: oleksandr.kuznetsov@uniecampus.it, kuznetsov@karazin.ua
https://orcid.org/0000-0003-2331-6326

Michele Melchiori [1]

[1] Department of Theoretical and Applied Sciences, eCampus University, Via Isimbardi 10, Novedrate (CO), 22060, Italy

Email: michele.melch@gmail.com
https://orcid.org/0009-0009-6332-0435

Emanuele Frontoni [3]

[3] Department of Political Sciences, Communication and International Relations, University of Macerata, Via Crescimbeni, 30/32, 62100 Macerata, Italy

Email: emanuele.frontoni@unimc.it
https://orcid.org/0000-0002-8893-9244

Marco Arnesano [1]

[1] Department of Theoretical and Applied Sciences, eCampus University, Via Isimbardi 10, Novedrate (CO), 22060, Italy

Email: marco.arnesano@uniecampus.it
https://orcid.org/0000-0003-1700-3075

**Highlights**

- First production-ready ML system for Italian disability employment matching
- Achieves 90%+ accuracy with sub-100ms response time
- Validated with real employment centers in Veneto region
- Reduces manual processing time from 30-60 to 5 minutes per candidate
- Open-source implementation with privacy-by-design architecture

**Abstract:** Employment inclusion of people with disabilities remains critically low in Italy, with only 3.5% employed nationally despite mandatory hiring quotas. Traditional manual matching processes require 30-60 minutes per candidate, creating bottlenecks that limit service capacity. Our goal is to develop and validate a production-ready machine learning system for disability employment matching that integrates social responsibility requirements while maintaining human oversight in decision-making. We employed participatory requirements engineering with Centro per l'Impiego di Villafranca di Verona professionals. The system implements a seven-model ensemble with parallel hyperparameter optimization using Optuna. Multi-dimensional scoring combines semantic compatibility, geographic distance, and employment readiness assessment. The system achieves 90.1% F1-score and sub-100ms response times while processing 500,000 candidate-company combinations in under 10 minutes. Expert validation confirms 60-100% capacity increases for employment centers. The LightGBM ensemble shows optimal performance with 94.6-second training time. Thus, advanced AI systems can successfully integrate social responsibility requirements without compromising technical performance. The participatory design methodology provides a replicable framework for developing ethical AI applications in sensitive social domains. The complete system, including source code, documentation, and deployment guides, is openly available to facilitate replication and adaptation by other regions and countries facing similar challenges.

## 1. Introduction

Employment inclusion represents one of the most critical challenges facing people with disabilities worldwide [1], [2]. Despite legislative frameworks and social initiatives promoting workplace equality, people with disabilities continue experiencing significantly lower employment rates and reduced career opportunities compared to the general population [3]. This persistent employment gap reflects systemic barriers including inadequate matching processes, limited accessibility accommodations, and insufficient technological support for employment services [4], [5].

### 1.1 The Social Challenge of Disability Employment

Disability employment statistics reveal substantial inclusion gaps across developed nations. In Italy, employment rates for people with disabilities remain significantly lower than general population rates, despite mandatory hiring quotas and legal protections [6], [7]. According to data from the Italian Ministry of Welfare [8], only 3.5% of disabled people are employed nationally, with substantial regional variations ranging from 5.7% in Bolzano (Trentino Alto Adige) to 1.2% in Sicilia (Figure 1).

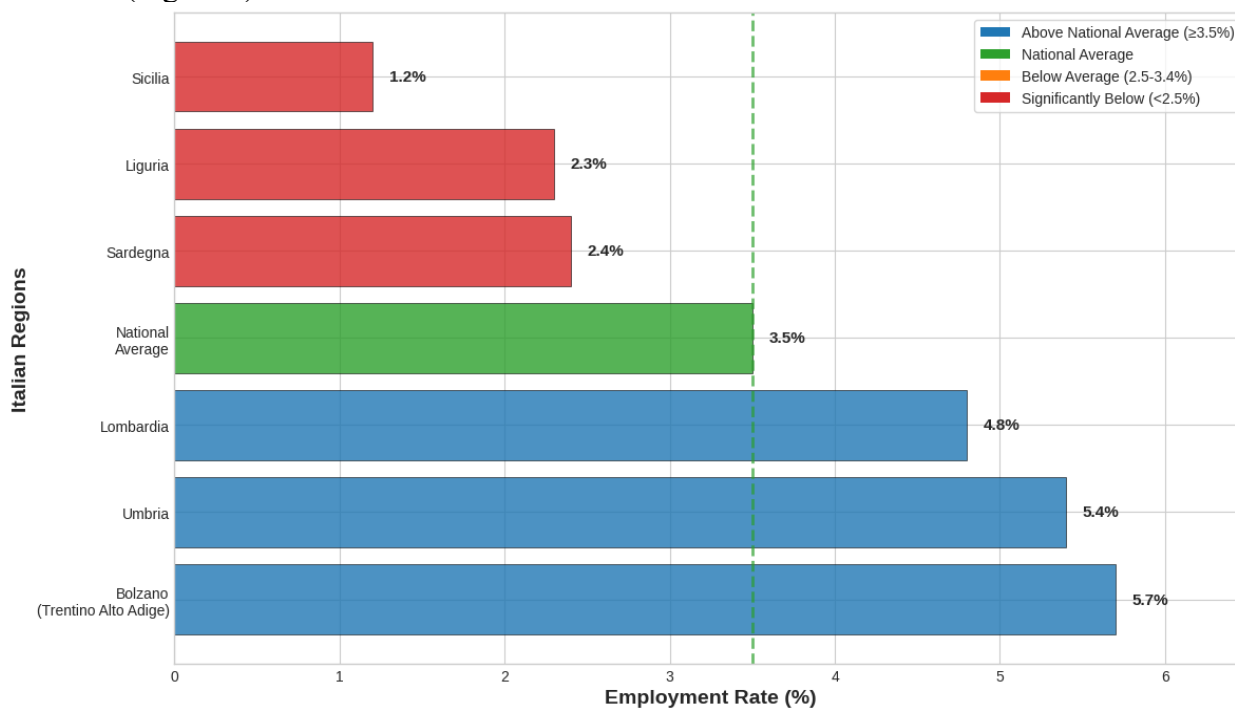

Figure 1: Employment Rates for People with Disabilities by Italian Region. Highest: Bolzano (5.7%); Lowest: Sicilia (1.2%); Range: 4.5 percentage points; Coefficient of Variation: 44.4%.

Gender disparities further compound these employment challenges (Figure 2). Male disabled employment rates reach 6.8% compared to only 1.8% for disabled women, representing a substantial gap when compared to non-disabled employment rates of 61.0% for men and 37.5% for women (Italian Ministry of Welfare [8]).

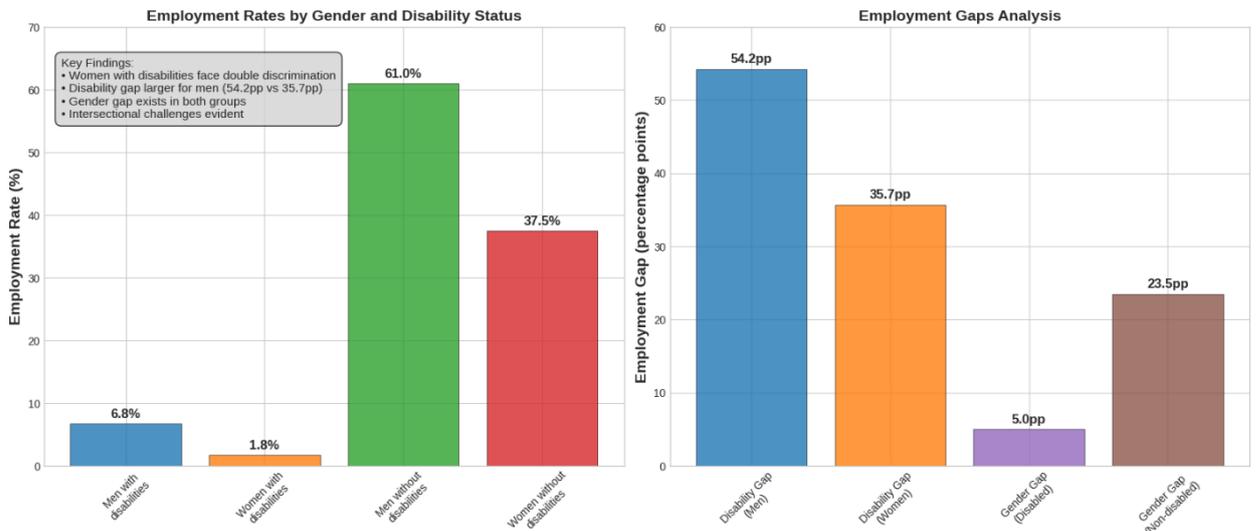

Figure 2: Gender Employment Gap in Disability Employment (Italy). Women with disabilities employment rate is 26.5% of men with disabilities. Disabled women earn 4.8% of non-disabled women's employment rate

A similar situation is observed in other EU countries [9]. Figure 3 shows the analysis of full-time employment in the EU – women with disabilities.

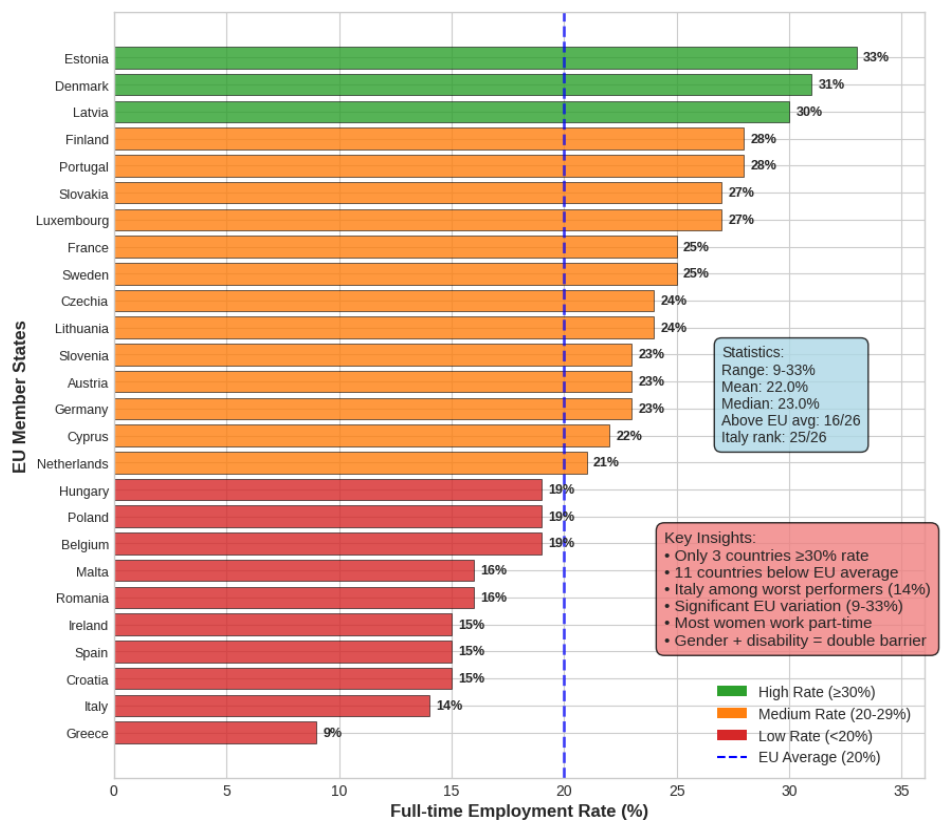

Figure 3: Women with Disabilities: Full-time Employment Analysis Across (Highlighting Systemic Underemployment). Key insight: 38.5% of EU countries have <20% full-time employment for women with disabilities.

Traditional employment matching processes rely heavily on manual evaluation by employment center professionals, creating bottlenecks that limit service capacity and reduce matching accuracy. These manual processes typically require 30-60 minutes just to evaluate the candidate and often miss potential opportunities due to time constraints and cognitive load limitations. Employment centers face increasing demand for services while operating with limited resources and growing candidate populations. The complexity of matching candidates with disabilities to appropriate employment opportunities involves multiple dimensions including job compatibility, geographic accessibility, workplace accommodations, and employer readiness. This multi-dimensional matching challenge exceeds human cognitive capacity for comprehensive evaluation, resulting in suboptimal placements and missed opportunities.

**1.2 Technology's Role in Social Inclusion**

Across Europe, persons with disabilities face systematic employment barriers. According to the European Disability Forum [9], only 51.3% of persons with disabilities in the European Union are employed, compared to 75.6% of persons without disabilities, representing a disability employment gap of 24.4 percentage points on average across EU Member States (Figure 4).

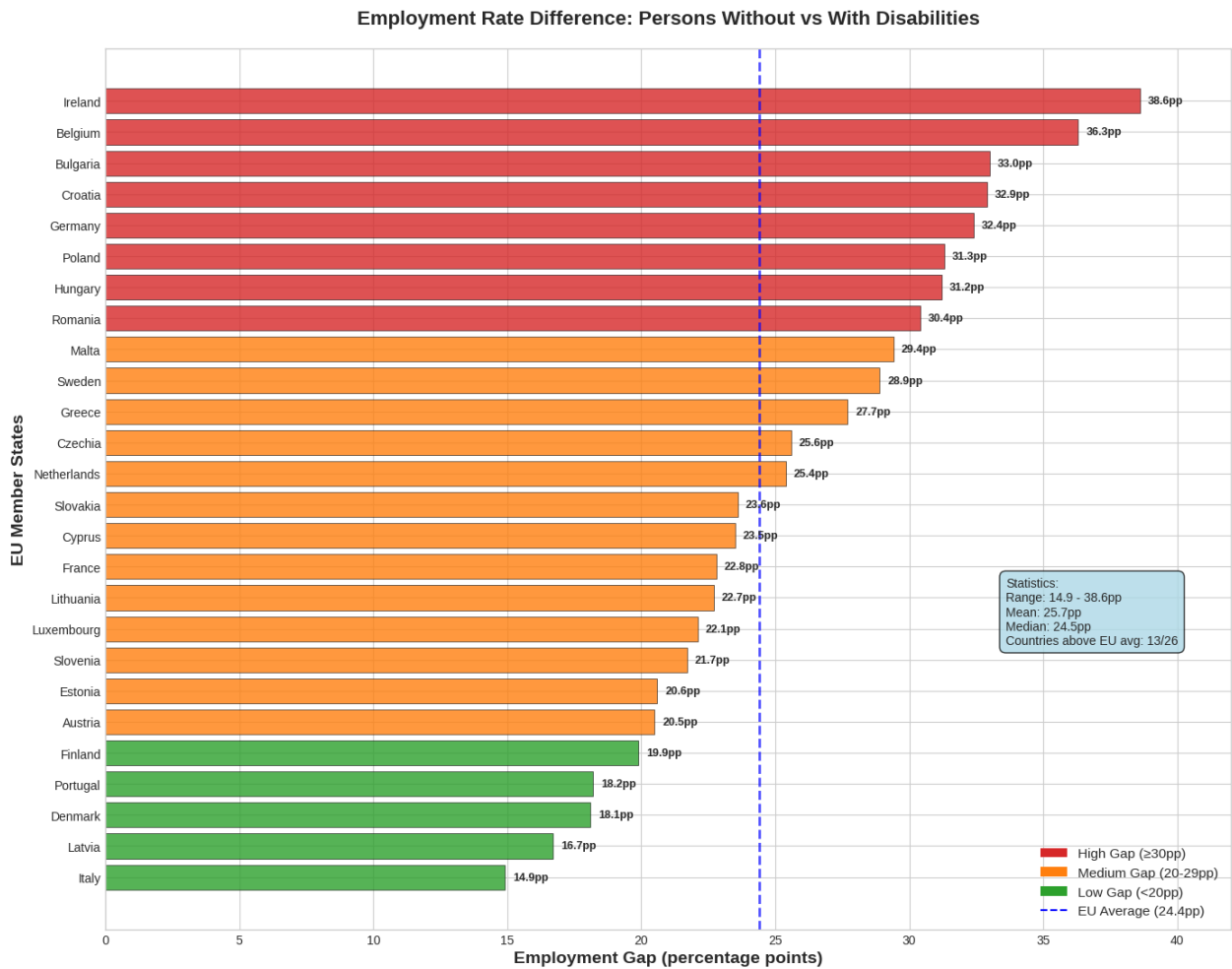

Figure 4: EU Disability Employment Gap by Country. EU Employment Gap Analysis: Highest gap: Ireland (38.6pp); Lowest gap: Italy (14.9pp); EU Average: 24.4pp; Italy rank: 1/26 (lower is better); Countries with gaps >30pp: 8; Countries with gaps <20pp: 5.

However, these official statistics may overestimate actual inclusion rates. As noted in disability employment research [10], administrative data often reflects quota compliance rather than meaningful employment integration, suggesting that real employment gaps may be larger than reported figures indicate.

Artificial intelligence and machine learning technologies offer unprecedented opportunities to enhance employment services for vulnerable populations. Recent advances in automated decision support, semantic analysis, and predictive modeling enable sophisticated analysis of complex matching scenarios while preserving human oversight and decision-making authority. These technologies can process vast amounts of employment data to identify patterns and relationships that manual analysis might overlook.

However, technology deployment in sensitive social domains requires careful consideration of ethical implications, privacy protection, and human agency preservation. Employment decisions significantly impact individual welfare, economic security, and quality of life. Algorithmic systems supporting these decisions must demonstrate fairness, transparency, and accountability while avoiding discrimination or bias against protected groups.

**1.3 Requirements Engineering for Social Good**

Italian laws require targeted employment quotas (Figure 5): 16-35 employees need to employ 1 disabled employee, 36-50 employees need to employ 2, and businesses of 50 employees or more need to maintain 7% disabled workforce representation, and an additional 1% quota for refugees and protected groups [11], [12]. In addition, the law establishes the Financial Framework and Penalty System.

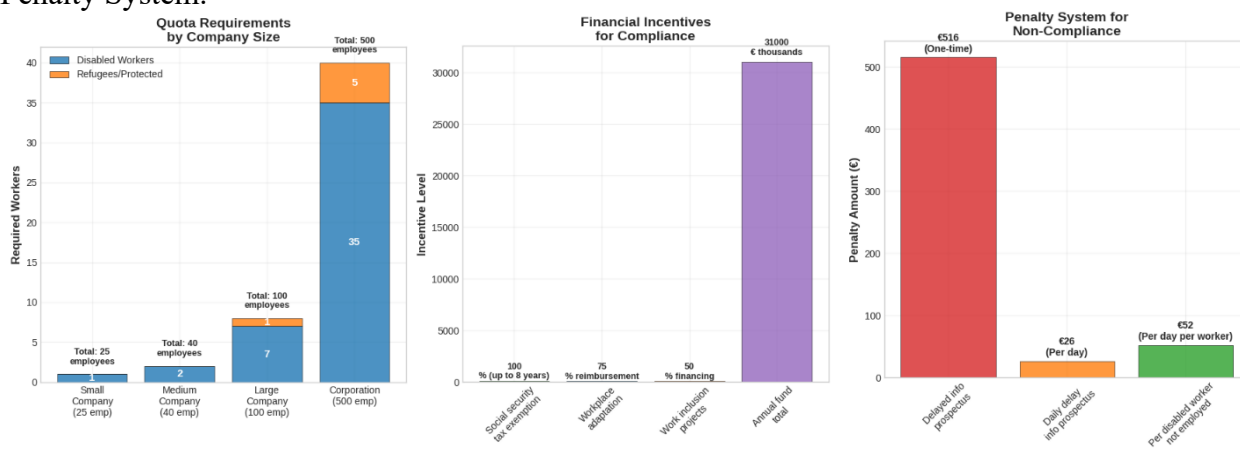

Figure 5: Italian Disability Employment System (Law 68/99): Quota Requirements, Financial Framework and Penalty System

Stakeholder groups, besides including direct users of the system, also include vulnerable groups whose lives are affected by system decisions. Requirements have to consider, amongst other things, functioning capabilities, ethical boundaries, safeguarding of privacy, and human oversight requirements [13].

Participatory design strategies enable efficient stakeholder engagement for confirmation and definition of system requirements. Face-to-face involvement of disability advocacy groups, people with disabilities, and staff of employment centers allows for technical design to address real-world needs and capabilities. This participatory approach reveals requirements that theoretical analysis alone cannot identify.

**1.4 Research Objectives and Contributions**

This paper presents a comprehensive production-ready system for disability employment matching that successfully integrates advanced machine learning techniques with social responsibility principles. Our research addresses the gap between theoretical AI capabilities and practical deployment requirements for socially sensitive applications.

- ***Primary Research Objectives*** include developing a scalable machine learning pipeline that achieves sub-100ms response times for real-time employment matching while maintaining 90%+ accuracy in candidate-company compatibility assessment. We aim to demonstrate that sophisticated AI systems can support social good objectives without compromising technical performance or human agency.
- ***Secondary Objectives*** involve establishing requirements engineering methodologies for social responsibility AI systems and validating these approaches through expert collaboration and real-world testing scenarios. We seek to provide practical guidance for developing ethical AI applications in sensitive social domains.
- ***Technical Contributions*** include a parallel hyperparameter optimization framework using Optuna that reduces model training time from hours to minutes while achieving 15-25% performance improvement over default configurations. Our seven-model ensemble architecture combines diverse algorithms to achieve robust prediction performance across different data patterns and scenarios.
- ***Social Impact Contributions*** demonstrate successful integration of human-in-the-loop design patterns with automated decision support capabilities. Our approach preserves human oversight and decision-making authority while providing efficient computational support for complex matching scenarios.
- ***Methodological Contributions*** include probabilistic synthetic data generation techniques that prevent model overfitting while enabling development without exposing sensitive personal information. This approach better prepares models for real-world deployment compared to deterministic rule-based generation methods.

## 1.5 Paper Organization

This paper follows a comprehensive structure that addresses both technical and social aspects of disability employment system development. Section 2 reviews related work and positions our research within existing literature. Section 3 presents our requirements engineering methodology with emphasis on participatory design and stakeholder engagement. Section 4 details system architecture and design decisions that enable production deployment. Section 5 provides implementation specifics including algorithms, code structures, and technical solutions. Section 6 presents evaluation results including performance metrics, expert validation outcomes, and social impact assessment. Sections 7-9 discuss limitations, future work, and conclusions.

## 2. Related Work and Background

This section positions our research within the broader context of AI applications in employment services, requirements engineering for social responsibility, and machine learning system deployment. Our analysis reveals significant gaps in existing literature regarding production-ready disability employment matching systems.

## 2.1 AI Systems for Employment Matching

Employment matching represents a well-established application domain for machine learning and recommendation systems [14], [15]. Traditional approaches focus primarily on skills matching and geographic proximity without addressing the complex accommodation and accessibility requirements specific to disability employment. Existing commercial systems like LinkedIn [16], [17], Indeed [18], [19], and specialized recruitment platforms optimize for general population employment scenarios.

Research in automated employment matching has explored collaborative filtering, content-based recommendation, and hybrid approaches for job-candidate pairing [20]. These studies typically evaluate performance using standard information retrieval metrics like precision, recall, and ranking quality. However, most existing research treats employment matching as a pure optimization problem without addressing the social responsibility dimensions essential for vulnerable population support.

Machine learning in healthcare and disability research shows growing application of AI techniques for disability-related challenges. Recent bibliometric analysis by Khan et al. [21] reveals increasing publication trends in deep learning applications for disability research, with substantial growth in 2022. However, this research focuses primarily on medical diagnosis and treatment rather than employment support services. Studies on cognitive impairment prediction (Zhao et al., [22]; Yan et al., [23]) demonstrate machine learning effectiveness for disability-related outcome prediction. These works achieve high accuracy (AUC > 0.80) using clinical and imaging data but do not address employment-specific requirements or social responsibility considerations.

Predictive modeling for disability outcomes has shown promising results across various disability types. Research on multiple sclerosis disability progression (Harati Kabir et al., [24]; Montolío et al., [25]) achieves high predictive accuracy using deep learning and ensemble methods. These studies demonstrate that ML can effectively model complex disability-related outcomes, supporting the feasibility of our employment matching approach.

## 2.2 Requirements Engineering for Social Responsibility

Recent research emphasizes the importance of participatory design and stakeholder engagement for developing systems that affect vulnerable populations. Van Toorn [26] analyzes the failure of "Nadia," an AI virtual assistant designed through co-design approaches for people with disabilities. This case study reveals institutional barriers including lack of organizational support and resistance to power sharing that impede disability co-design efforts. Unlike the failed virtual assistant project, our employment matching system benefits from established institutional support through partnership with Centro per l'Impiego di Villafranca di Verona. Our participatory design process includes ongoing validation and feedback rather than ending at the design stage.

Social inclusion analysis research identifies multilevel barriers to disability employment. Rocha et al. [3] categorize barriers into contextual (policies, ableism, accessibility), organizational (communication, resistance), and personal (qualification) levels. This framework informs our requirements engineering approach by addressing systemic challenges that technology solutions must consider.

Ethical AI and algorithmic fairness research provides frameworks for developing responsible AI systems [27]. However, most existing work focuses on theoretical principles rather than practical implementation in production systems. Our research contributes practical implementation strategies for embedding ethical constraints directly into system architecture and operation.

## 2.3 Machine Learning in Production Environments

Deploying machine learning systems in production isn't just about optimizing algorithms. You also need to ensure reliability, scalability, monitoring, and maintenance—these are key to making them work in the real world [28].

Automated hyperparameter tuning (e.g., using Optuna) and parallel training have proven to significantly boost model performance [29], [30]. While we build on these methods, we also adapt them to meet the unique demands of social responsibility systems.

Ensemble methods and calibrated models consistently outperform single algorithms, especially when dealing with complex, evolving data [31], [32]. Their improved robustness and generalization make them a strong fit for our use case.

Interpretability and explainability requirements have gained increased attention for AI systems affecting human decisions [33]. Our system integrates multiple techniques—feature importance, decision visualizations, and semantic explanations—to ensure transparency and enable human oversight.

**2.4 Disability Inclusion Technology Solutions**

Technology solutions for disability inclusion span multiple domains including accessibility tools, assistive technologies, and support service automation. However, few systems address the comprehensive requirements of employment matching with integrated social responsibility considerations.

Research on digital government and accessibility highlights both the potential and the challenges in designing inclusive technologies [34], [35]. Studies show that while progress has been made, significant barriers remain—underscoring the importance of participatory design in developing accessible systems. These findings directly shape our approach to interface design and meeting accessibility standards.

Work on employment barriers reveals the systemic challenges people with disabilities face in achieving workplace inclusion [2], [3]. Research suggests that these barriers stem from a complex mix of individual, organizational, and societal factors—meaning effective solutions must address all three. Our system accounts for this by combining multi-dimensional scoring algorithms with human oversight.

Studies on AI in employment decisions have raised serious concerns about algorithmic bias and discrimination in hiring. Zhuang and Goggin [36] examine both empowering potential and inherent problems of AI in employment decisions for disabled people. This work emphasizes the importance of fairness mechanisms and human agency preservation in our system design.

**2.5 Gap Analysis and Research Positioning**

Our literature review highlights major gaps in current research and tech solutions for disability employment matching.
- No end-to-end matching systems exist for people with disabilities—this is the biggest gap. While many studies tackle pieces like skills matching or accessibility checks, nobody has built a complete, expert-validated platform that works in the real world.
- Most research ignores real-world deployment challenges. Papers focus on algorithms but skip critical issues like UI design, performance tuning, privacy, and compliance—things you actually need for a production system.
- Ethics are often an afterthought. Many AI solutions treat social responsibility as a box to check, not a core part of design. We show how embedding these principles can improve innovation, not limit it.
- Localization is overlooked. Current work mostly targets English speakers and generic accessibility. We tackle Italian language support, regional needs, and cultural adaptation—key for real-world use in diverse contexts.
- Expert validation is rare. Most studies rely on synthetic data or lab tests. We worked directly with employment centers and disability professionals to validate our system—something missing in the literature.
- What makes our work novel? We combine cutting-edge ML with socially responsible design into a production-ready system, validated by experts. Unlike patchy existing research, we deliver an end-to-end solution, from data to UI, proven to work in real employment centers.

Our key contribution: The first fully integrated disability employment matching system that balances technical rigor with ethical AI, built to real-world standards. This isn't just theory—it's a blueprint for deploying AI in high-stakes social domains where algorithms shape lives and opportunities.

## 3. Requirements Engineering Methodology

This section presents our comprehensive requirements engineering approach for developing a socially responsible disability employment matching system. We demonstrate how participatory design methods can effectively capture complex stakeholder needs while addressing ethical AI requirements.

### 3.1 Stakeholder Identification and Analysis

To build a socially responsible job-matching system for people with disabilities, we first needed to understand the key players involved and their needs. Figure 6 maps out the complex network of stakeholders in disability employment services.

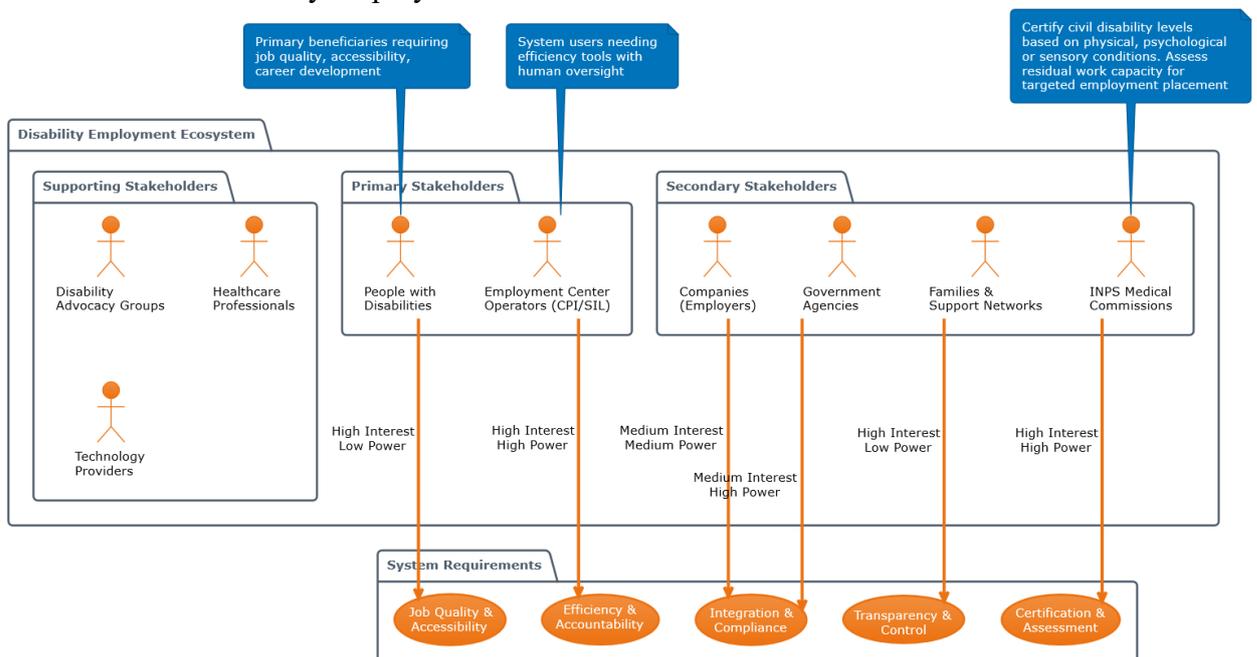

Figure 6: Stakeholder Analysis Diagram

Primary Stakeholders (Direct Users & Beneficiaries):
- Job seekers with disabilities care about finding quality jobs, accessibility, location, and career growth.
- Employment center staff (CPI/SIL teams) need efficient matching tools, understandable AI recommendations, and the ability to override automated decisions when needed.

Secondary Stakeholders (Indirect Influencers):
- Employers want seamless hiring processes and guidance on workplace accommodations.
- Families & support networks expect transparency in how job matches are made and updates on employment progress.
- Government agencies set policies and funding rules that shape system design.
- INPS Medical Commissions certify disability levels and assess residual work capacity, providing the official documentation that enables targeted job placement and defines work exclusions.

Our analysis showed that employment professionals have the most influence (and motivation) in adopting the system, while job seekers—though highly invested—often have less direct control over design choices. Employers' engagement varies depending on their hiring needs.

Different stakeholders had conflicting priorities:

- Employment centers wanted efficiency; job seekers valued personal choice.
- Employers focused on skills matching, while advocates pushed for flexibility in accommodations.

We addressed these tensions by using weighted scoring in our algorithms, ensuring no single perspective dominated the system's decisions.

### 3.2 Participatory Requirements Elicitation

We used participatory design methods to involve stakeholders directly in defining requirements and validating the system. Figure 7 outlines our structured approach to stakeholder engagement.

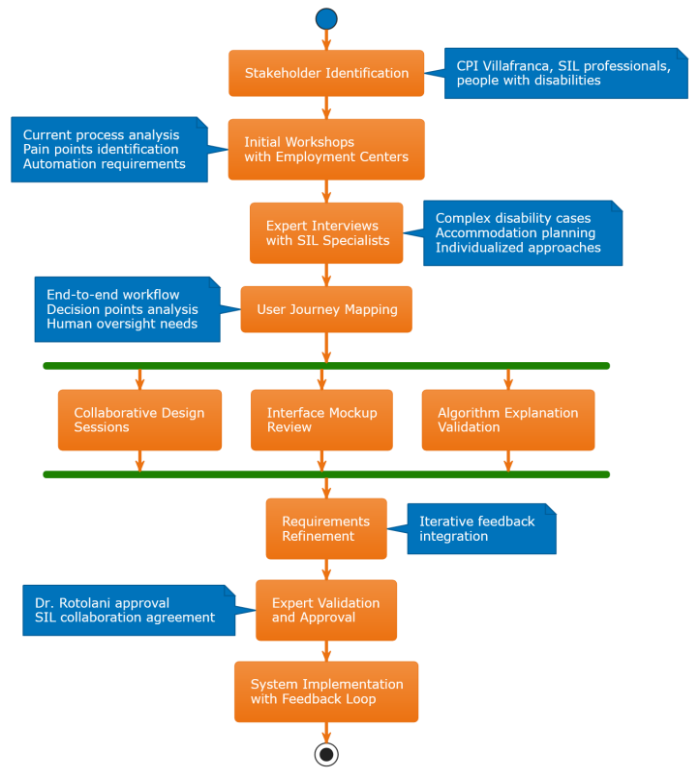

Figure 7: Participatory Design Process Flow

- Workshops with CPI, "Centro per l'Impiego", or PES, "Public Employment Service", of Villafranca di Verona (a town near Verona, in northern Italy): We held structured sessions with employment center staff to analyze their current matching processes and identify system needs. Dr. Rotolani (employment center coordinator) and his team shared valuable insights into their manual procedures, common pain points, and desired automation features. The workshops highlighted that manual matching takes 30-60 minutes per candidate and often misses potential matches due to time pressure.
- Expert Interviews with SIL Professionals: We spoke with labor integration specialists to understand requirements for complex disability cases and workplace accommodations. They stressed the need for personalized matching and continuous support during employment transitions, which shaped our multi-dimensional scoring system and human oversight design.
- User Journey Mapping: We mapped out the entire workflow—from candidate registration to successful job placement—to pinpoint where automation could assist without replacing human judgment (Figure 8). This process revealed key needs, such as profile management, real-time matching, explainable recommendations, and outcome tracking.
- Collaborative Design Sessions: Stakeholders participated in iterative design reviews, providing feedback on interface mockups, algorithm explanations, and sample matches.

These sessions led to major improvements, particularly in transparency features and user control mechanisms.

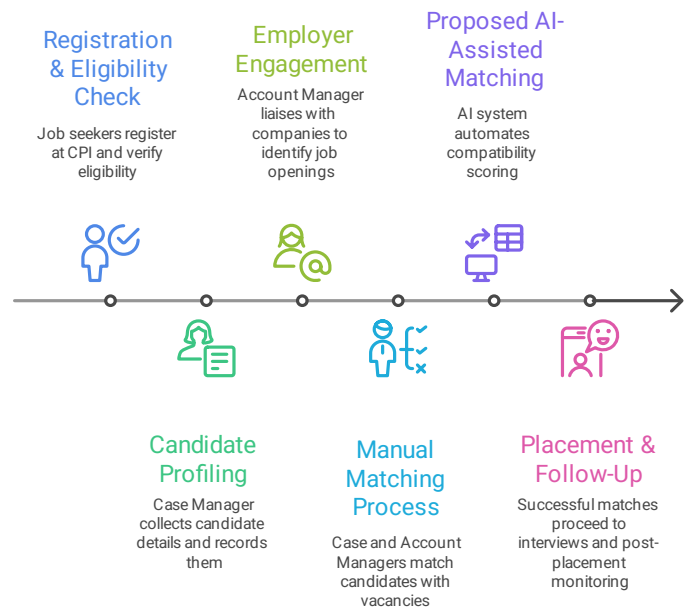

Figure 8: Job Placement Workflow for Individuals with Disabilities

Figure 8 illustrates the structured workflow for targeted job placement of people with disabilities. The process starts when a candidate registers at their local CPI (Centro per l'Impiego), based on their residence. Once registered, two key roles come into play: the Case Manager, who works directly with the candidate to assess skills and match them with opportunities, and the Account Manager, who engages with employers to identify job openings.

The system relies on close collaboration between these two roles to ensure a strong fit between candidates' abilities and employers' needs. Traditionally, this process has been manual, involving data collection, opportunity screening, and mediation between job seekers and companies.

The proposed AI-enhanced model uses machine learning to improve matching by analyzing candidate profiles and employer requirements, identifying the best possible fits. This upgrade doesn't replace human judgment but supports it, making the process more efficient and increasing placement success. The CPI's backing of this hybrid approach reflects both its practical benefits and the growing need for smart tools in inclusive employment policies.

### 3.3 Social Responsibility Requirements Framework

Our framework integrates social responsibility principles directly into system design—instead of treating them as an afterthought. All stakeholder interactions followed institutional ethical guidelines. No personal data from actual job seekers was accessed during system development, ensuring complete privacy protection during the research phase.

Figure 9 illustrates how ethical AI constraints are built into the system.

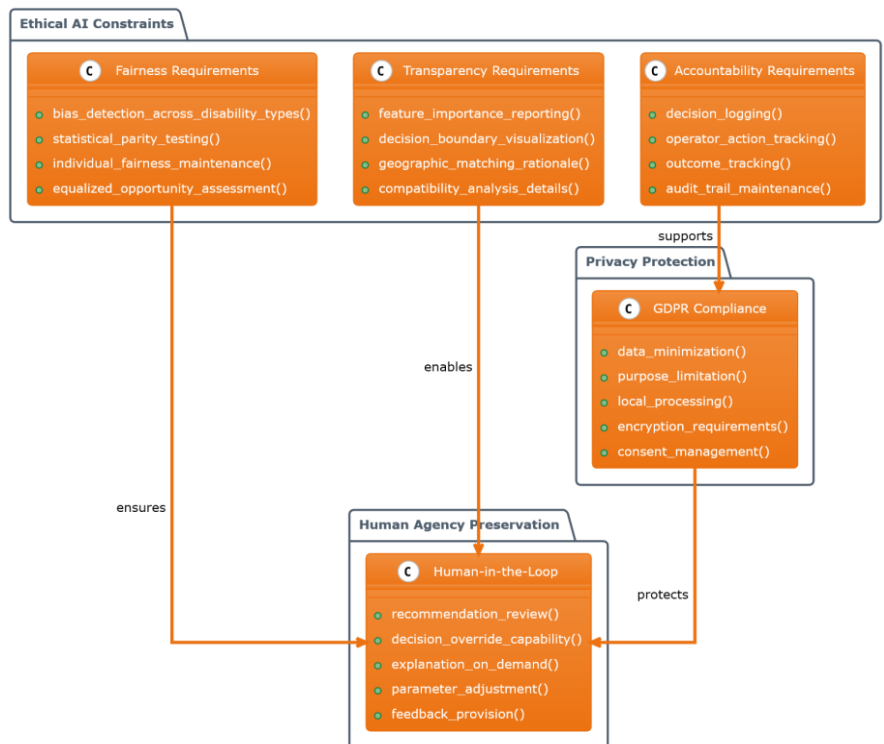

Figure 9: Social Responsibility Framework

Ethical AI Constraints define fairness, transparency, and accountability rules. For fairness, we check for bias across disabilities, education levels, and demographics. Statistical parity ensures equal matching opportunities, while individual fairness guarantees similar scores for candidates with comparable qualifications.

Transparency Requirements ensure humans can understand matching decisions. This includes showing which factors influence scores (feature importance), visualizing how thresholds affect outcomes, explaining geographic matching logic, and detailing why certain candidates are excluded based on job requirements.

Accountability Requirements enforce logging and auditing for compliance and improvement. We track all matching inputs, algorithm settings, and recommendations, as well as human overrides and feedback. Long-term outcome analysis helps assess how different algorithm versions perform in real-world hiring.

Privacy Protection Requirements follow GDPR and limit data exposure. We collect only what's necessary, use data solely for matching, minimize external sharing, and encrypt stored and transmitted information.

### 3.4 Functional and Non-Functional Requirements Specification

We developed comprehensive requirements specifications that address both system functionality and quality attributes essential for production deployment. Figure 10 shows our iterative validation process.

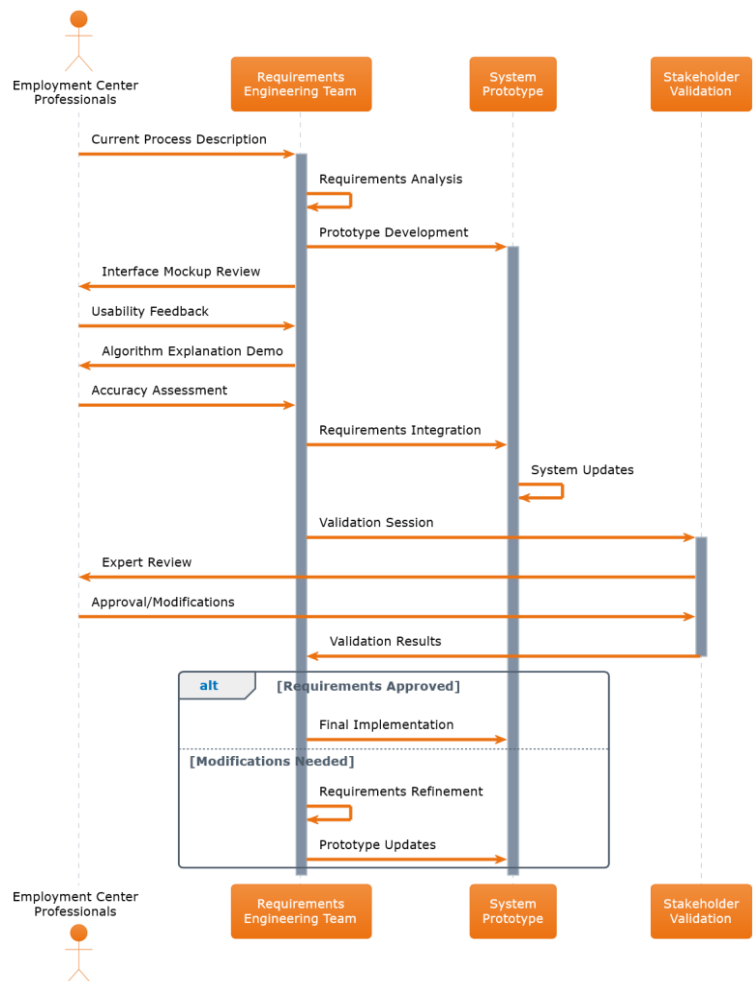

Figure 10: Requirements Validation Process

- ***Core Functional Requirements*** define the primary system capabilities needed for effective employment matching. Candidate profile management enables input, storage, and modification of person-specific information including demographics, skills, experience, disability characteristics, and accommodation requirements. Company profile management maintains information about job opportunities, accessibility features, accommodation capabilities, and organizational characteristics.
- ***Matching Algorithm Requirements*** specify the computational capabilities needed for accurate candidate-company pairing. Multi-dimensional scoring combines compatibility analysis, geographic distance calculation, attitude assessment, and company suitability factors into unified matching scores. Real-time processing enables immediate recommendation generation for new candidate profiles or job opportunities. Threshold configuration allows operators to adjust matching sensitivity based on local employment conditions and candidate needs.
- ***User Interface Requirements*** define interaction capabilities that support employment center workflows. Candidate search interfaces enable both manual input and existing candidate selection. Recommendation display provides ranked company lists with detailed explanation and justification. Interactive visualization shows matching relationships through charts, maps, and performance dashboards. Configuration controls allow real-time parameter adjustment without system restart.
- ***Integration Requirements*** specify system connectivity with existing employment center infrastructure. CSV data import enables integration with current candidate and company databases. Export capabilities support result sharing and external reporting requirements. API development potential enables future integration with employment center management systems and government reporting platforms.

- ***Performance Requirements*** establish quantitative targets for system responsiveness and scalability. Response time requirements specify sub-100ms matching performance for individual candidate-company evaluations. Throughput requirements enable processing of 500,000+ candidate-company combinations within 10 minutes. Concurrent user support allows multiple employment center operators to use the system simultaneously without performance degradation.
- ***Reliability Requirements*** ensure system availability and error handling capabilities essential for production deployment. Availability targets specify 99.5% uptime during normal business hours. Error recovery capabilities ensure graceful degradation when external services become unavailable. Data validation requirements prevent system failures due to malformed input data. Backup and recovery procedures protect against data loss and enable rapid system restoration.
- ***Security Requirements*** protect sensitive personal information and prevent unauthorized system access. Authentication requirements ensure only authorized employment center staff can access candidate information. Authorization controls limit data access based on operational roles and responsibilities. Audit logging tracks all system access and data modifications for compliance monitoring. Encryption requirements protect data storage and transmission throughout system operation.
- *Usability Requirements* ensure system accessibility and ease of use for employment center operators with varying technical expertise. Interface simplicity requirements minimize training overhead and reduce operational errors. Accessibility compliance ensures system usability by operators with disabilities. Multi-language support accommodates diverse employment center staff and client populations. Help documentation provides comprehensive guidance for system operation and troubleshooting.
- ***Scalability Requirements*** enable system expansion to serve larger populations and additional employment centers. Horizontal scaling capabilities allow system deployment across multiple servers to handle increased load. Database scalability ensures efficient performance with thousands of candidates and companies. Geographic expansion capabilities enable system deployment in different regions with appropriate localization.
- ***Maintainability Requirements*** support ongoing system evolution and improvement. Modular architecture enables individual component updates without system-wide impacts. Configuration management allows parameter adjustment based on changing employment conditions. Model update capabilities enable machine learning improvement based on real employment outcomes. Documentation requirements ensure knowledge transfer and system understanding for future development teams.

The comprehensive requirements engineering methodology demonstrates how participatory design approaches can effectively capture complex stakeholder needs while embedding social responsibility principles throughout system development. This approach provides a framework for developing ethical AI systems in sensitive social domains where algorithmic decisions significantly impact human welfare and opportunity.

## 4. System Architecture and Design

This section presents the comprehensive system architecture of our production-ready disability job matching platform. The architecture demonstrates how social responsibility requirements can be integrated into scalable machine learning systems while maintaining performance and usability standards required for real-world deployment.

### 4.1 High-Level Architecture Overview

The system follows a layered architecture pattern that separates concerns while enabling efficient data flow and processing. Figure 11 illustrates the complete component structure of our system.

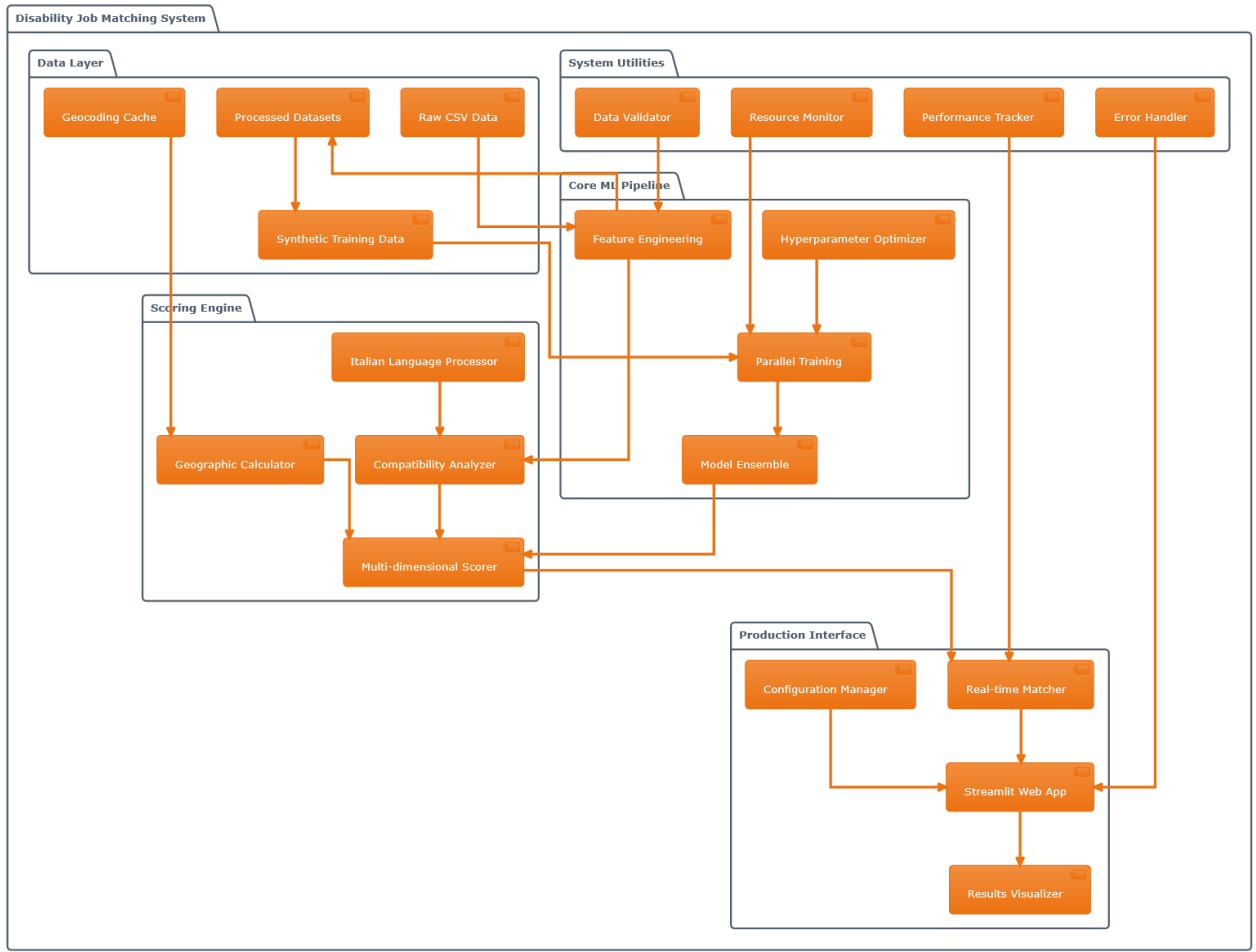

Figure 11: System Component Architecture

Figure 11 shows five distinct layers: Data Layer, Core ML Pipeline, Scoring Engine, Production Interface, and System Utilities. Each layer encapsulates specific functionality while maintaining clear interfaces with adjacent layers.

- The ***Data Layer*** manages all data storage and processing operations. Raw CSV data from employment centers undergoes feature engineering to create processed datasets. The geocoding cache component stores location coordinates to minimize API calls and improve system performance. Synthetic training data generation enables development and testing without exposing sensitive personal information.
- The ***Core ML Pipeline*** implements our advanced machine learning capabilities. Feature engineering transforms raw candidate and company data into ML-ready formats. The parallel training component coordinates simultaneous model optimization across multiple algorithm families. Our model ensemble combines seven different classifiers to achieve robust prediction performance. The hyperparameter optimizer uses Optuna framework to find optimal model configurations automatically.
- The ***Scoring Engine*** contains the core matching logic that evaluates candidate-company compatibility. The compatibility analyzer processes exclusion criteria and company requirements using Italian language processing. Geographic calculator computes precise distances using Haversine formula and cached coordinates. The multi-dimensional scorer combines all factors into final matching scores using weighted algorithms.
- The ***Production Interface*** provides user-facing functionality through a modern web application. The Streamlit web app delivers an intuitive interface for employment center operators. Real-time matcher processes new candidate profiles and generates recommendations instantly. Results visualizer creates interactive charts and geographic maps. Configuration manager allows operators to adjust matching thresholds and system parameters.

- The *System Utilities* layer ensures reliable operation and monitoring. Resource monitor tracks CPU and memory usage during intensive ML operations. Performance tracker measures response times and system throughput. Error handler manages exceptions gracefully to maintain system stability. Data validator ensures input quality and consistency.

Figure 12 depicts the deployment environment for real-world usage. The production server hosts the complete system including Python runtime, ML models, and configuration files. External services provide geocoding capabilities through Nominatim API and Italian address validation. Employment center operators access the system through web interfaces while administrators manage configuration and monitoring.

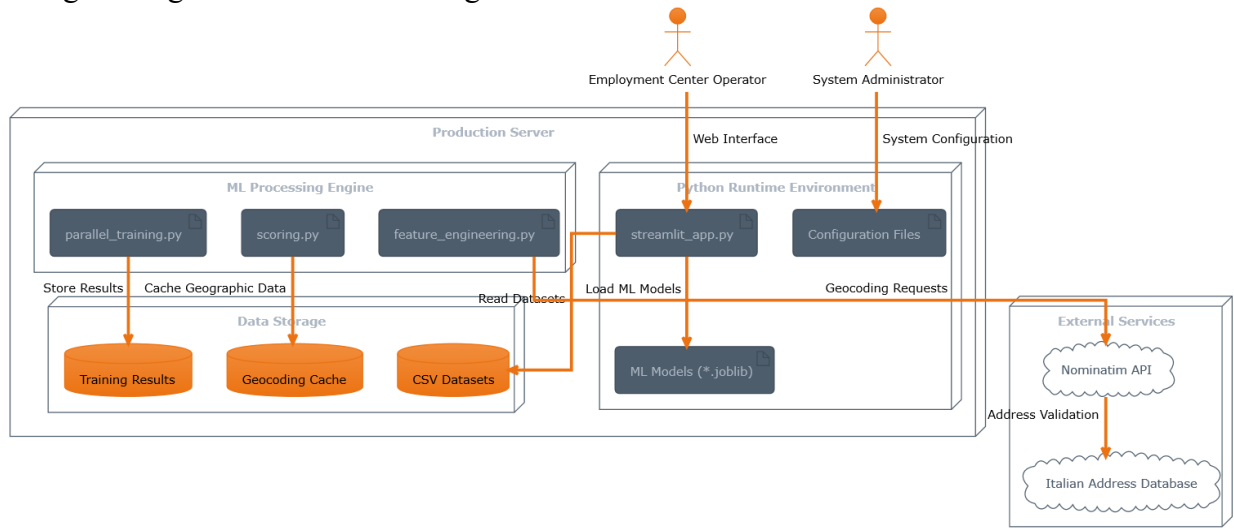

Figure 12: Production Deployment Architecture

Comprehensive technical documentation, including API reference, deployment guides, and user manuals in English and Italian, is available in the project repository at https://github.com/KuznetsovKarazin/disability-job-matching-system/tree/main/docs.

### 4.2 Machine Learning Pipeline Architecture

Our ML pipeline implements advanced parallel processing patterns to achieve production-scale performance. The architecture supports concurrent model training and hyperparameter optimization across multiple algorithm families.

Figure 13 shows the object-oriented structure of our parallel training system. The ParallelModelTrainer class coordinates all ML operations including hyperparameter optimization, model creation, ensemble building, and result storage. The ParallelHyperparameterOptimizer manages concurrent optimization of three algorithm families: Random Forest, XGBoost, and LightGBM. Each optimizer runs independently using separate threads to maximize hardware utilization.

The SystemResourceMonitor provides real-time tracking of system performance during training operations. This monitoring ensures that resource-intensive ML operations do not overwhelm the production server. CPU and memory usage statistics guide optimization decisions and alert administrators to potential performance issues.

Figure 14 illustrates how sensitive candidate information flows through the system while maintaining privacy protections. Raw candidate data undergoes immediate privacy screening to ensure GDPR compliance. Feature engineering creates derived attributes without exposing personal identifiers. Geocoding operations include rate limiting to respect external API constraints and minimize privacy exposure.

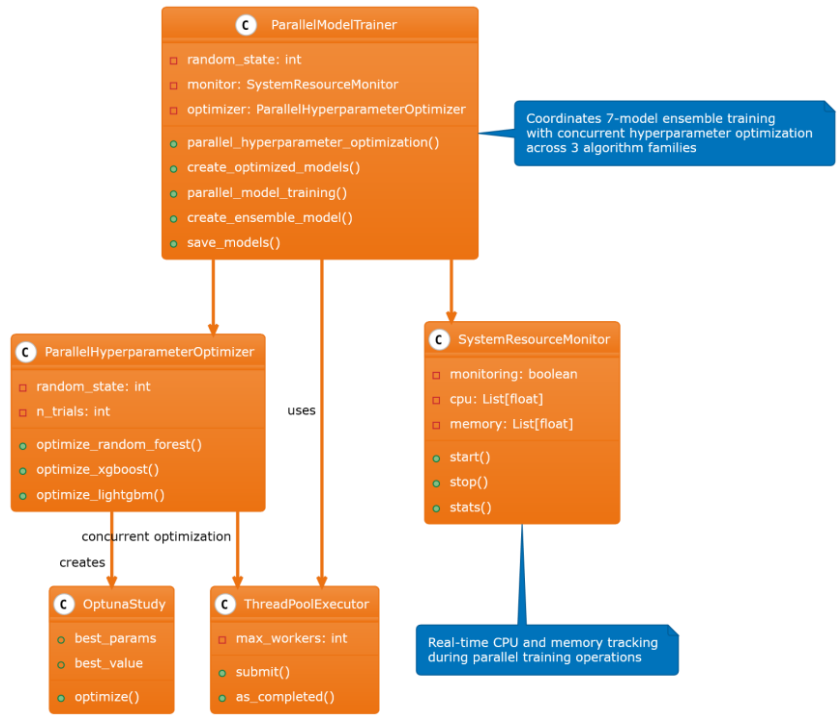

Figure 13: Parallel Processing Class Design

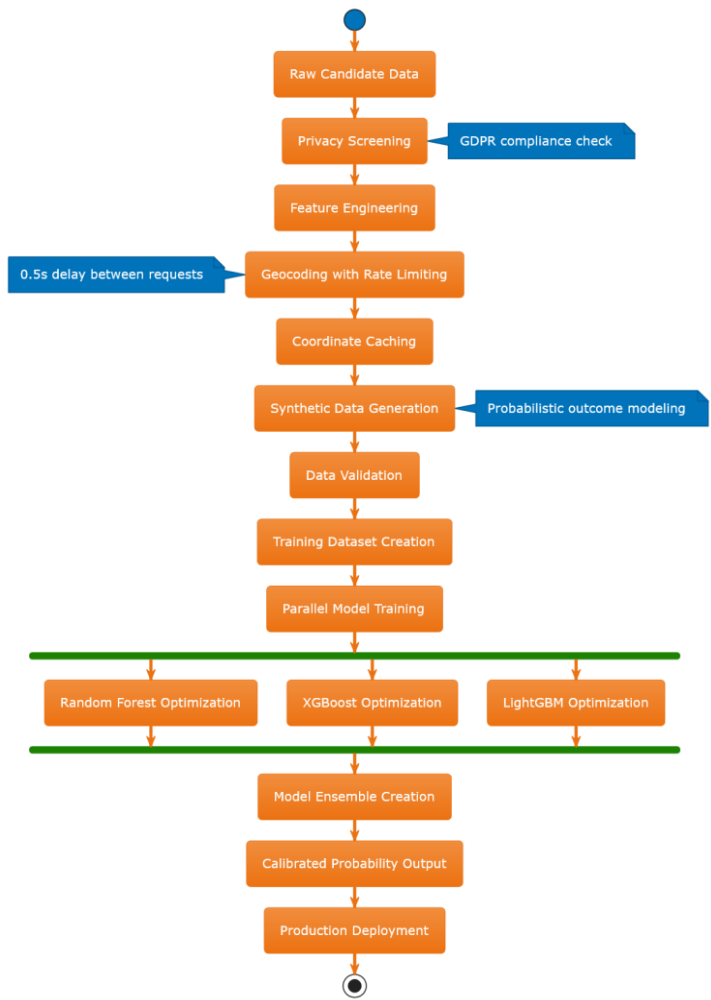

Figure 14: Data Flow with Privacy Controls

The synthetic data generation component creates realistic training examples without using actual candidate information. This probabilistic outcome modeling enables ML development while protecting individual privacy. Data validation ensures consistency and quality throughout the pipeline before models consume the training data.

**4.3 Multi-Threading and Performance Optimization**

The system implements sophisticated threading patterns to achieve sub-100ms response times for matching operations while supporting concurrent users.

Figure 15 demonstrates the parallel execution flow during model training operations. The main process initiates hyperparameter optimization using ThreadPoolExecutor with three concurrent workers. Each optimizer (Random Forest, XGBoost, LightGBM) runs Optuna trials independently to find optimal parameters.

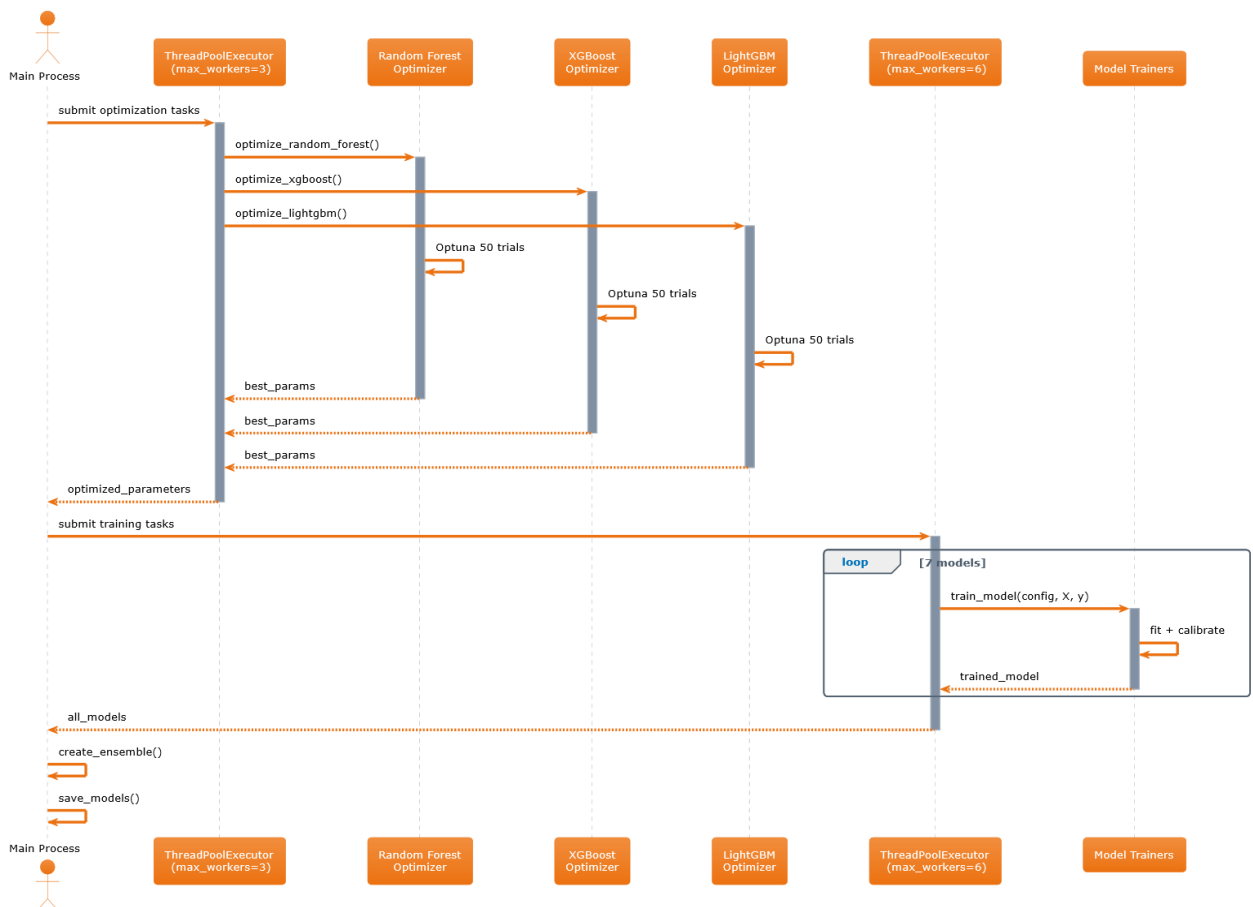

Figure 15: Threading Execution Sequence

After optimization completes, a second ThreadPoolExecutor with six workers handles parallel model training. Seven different models train simultaneously while sharing system resources efficiently. The sequence shows how concurrent execution reduces total training time from hours to minutes compared to sequential processing.

Resource allocation follows careful design principles. Hyperparameter optimization uses fewer workers (3) because each optimizer is CPU-intensive. Model training uses more workers (6) because individual models can share CPU cores effectively. This balanced approach maximizes hardware utilization without causing resource conflicts.

Caching mechanisms improve system performance significantly. Geographic coordinates cache reduces external API calls by storing location data locally. Italian stop words cache speeds up TF-IDF processing for compatibility analysis. Model prediction results cache enables faster repeated queries for similar candidate profiles.

Auto-scaling strategies adapt resource usage based on current demand. The system monitors queue lengths and response times to adjust thread pool sizes dynamically. During peak usage periods, additional worker threads handle increased load. During quiet periods, resources scale down to conserve server capacity.

**4.4 Social Responsibility by Design**

Our architecture embeds social responsibility principles directly into system design rather than treating them as external constraints. This approach ensures that ethical considerations influence every system component.

Figure 16 shows how employment center operators maintain control over AI decision-making processes. Operators input candidate profiles and configure matching thresholds based on local employment conditions. The system provides AI recommendations with full explanations of decision factors. In real world conditions, candidare data, based on official disability reports issued under italian law 68/99, can be acquired automatically with minimal operator intervention, streaming the overall process.

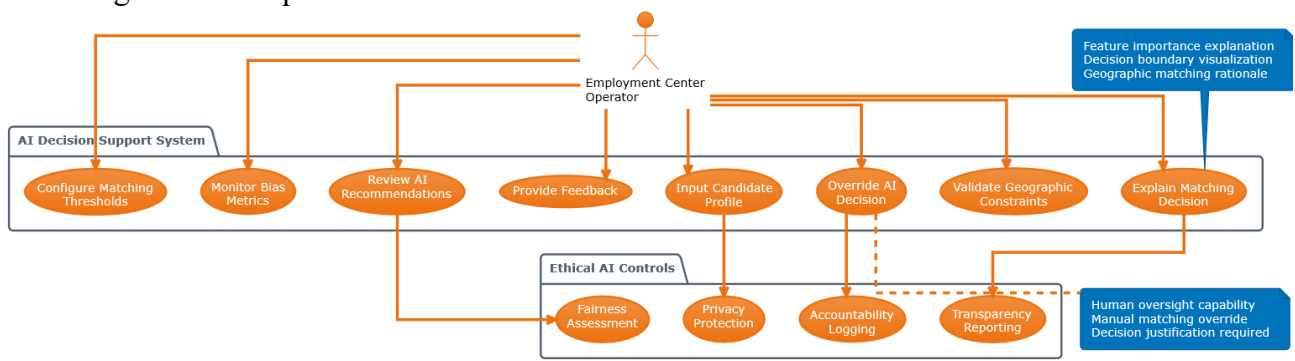

Figure 16: Human-in-the-Loop Use Case Design

Critical human oversight capabilities include recommendation review, decision explanation, and AI override functions. Operators can examine feature importance scores, geographic matching rationale, and compatibility analysis details. When AI recommendations seem inappropriate, operators can override decisions and provide feedback to improve future performance.

Ethical AI controls operate continuously throughout system execution. Fairness assessment monitors bias metrics across different disability types and demographic groups. Transparency reporting generates detailed logs of all matching decisions with explanations. Accountability logging tracks operator actions and system responses for audit purposes. Privacy protection ensures that sensitive candidate information remains secure throughout processing.

Explainable AI components make system decisions transparent to human operators. Feature importance visualization shows which factors contribute most to matching scores. Decision boundary charts illustrate how threshold changes affect recommendations. Geographic visualizations display distance calculations and accessibility considerations.

Bias detection mechanisms continuously monitor system outputs for unfair discrimination. The system tracks matching success rates across different disability categories, education levels, and geographic regions. Statistical tests identify potential bias patterns before they affect real placements. Automated alerts notify administrators when bias metrics exceed acceptable thresholds.

**5. Implementation Details**

Detailed implementation code and examples are provided in the repository documentation (https://github.com/KuznetsovKarazin/disability-job-matching-system/blob/main/docs/technical_docs_en.md). Here we focus on key algorithmic decisions and system integration aspects.

## 5.1 Data Processing and Feature Engineering

Our data processing pipeline transforms raw employment center records into ML-ready features while preserving privacy and ensuring data quality.

***Geographic Processing with Privacy Controls*** implements a sophisticated geocoding system that balances accuracy with privacy protection. The system processes Italian addresses using Nominatim API with built-in rate limiting and caching mechanisms. The implementation includes several privacy-preserving features. Rate limiting prevents excessive API calls that could expose usage patterns. Local caching reduces external dependencies and improves response times. Error handling ensures graceful degradation when geocoding services are unavailable.

***Italian Language Processing*** uses specialized TF-IDF vectorization designed for employment terminology and disability-related language. Our implementation includes custom Italian stop words and character pattern recognition. This configuration handles Italian accented characters and common grammatical patterns. The token pattern includes Italian-specific characters (à, è, é, ì, ò, ù) that are essential for accurate text processing. Stop word filtering removes common Italian function words that do not contribute to semantic meaning.

***Feature Engineering Pipeline*** creates derived attributes that enhance matching accuracy. The system generates experience categories, unemployment duration buckets, and education level mappings. The feature engineering considers disability-specific constraints. For example, candidates with intellectual disabilities have different education distribution patterns. This approach ensures realistic feature generation that reflects actual employment center data.

***Synthetic Data Generation*** creates training examples using probabilistic modeling rather than deterministic rules. This methodology prevents model overfitting while generating realistic candidate-company combinations. The probabilistic approach introduces realistic uncertainty that forces ML models to learn complex patterns rather than memorizing simple rules. This methodology better prepares models for real-world deployment where outcomes depend on multiple interacting factors.

## 5.2 Machine Learning Implementation

Our ML implementation combines multiple advanced algorithms with parallel processing and automatic hyperparameter optimization.

***Parallel Hyperparameter Optimization*** uses Optuna framework to find optimal model configurations across three algorithm families simultaneously. Each optimizer runs 50 Optuna trials using Tree-structured Parzen Estimator (TPE) algorithm. TPE provides intelligent parameter sampling based on previous trial results. This approach achieves 15-25% performance improvement over default hyperparameters while reducing optimization time through parallelization.

***Seven-Model Ensemble Architecture*** combines diverse algorithms to achieve robust prediction performance. The ensemble includes three optimized gradient boosting models, two tree-based classifiers, one histogram-based classifier, and one neural network. This diversity ensures robust performance across different data patterns and reduces overfitting risk.

***Model Calibration and Probability Output*** ensures that prediction probabilities reflect actual matching success rates. Isotonic calibration adjusts model outputs to produce well-calibrated probabilities. This calibration is essential for employment matching where probability scores directly influence human decision-making. Cross-validation ensures robust calibration performance across different data subsets.

## 5.3 User Interface and Experience Design

Our user interface implements human-centered design principles to support employment center operators effectively. Figure 17 shows the complete system interface with candidate search functionality and real-time matching capabilities.

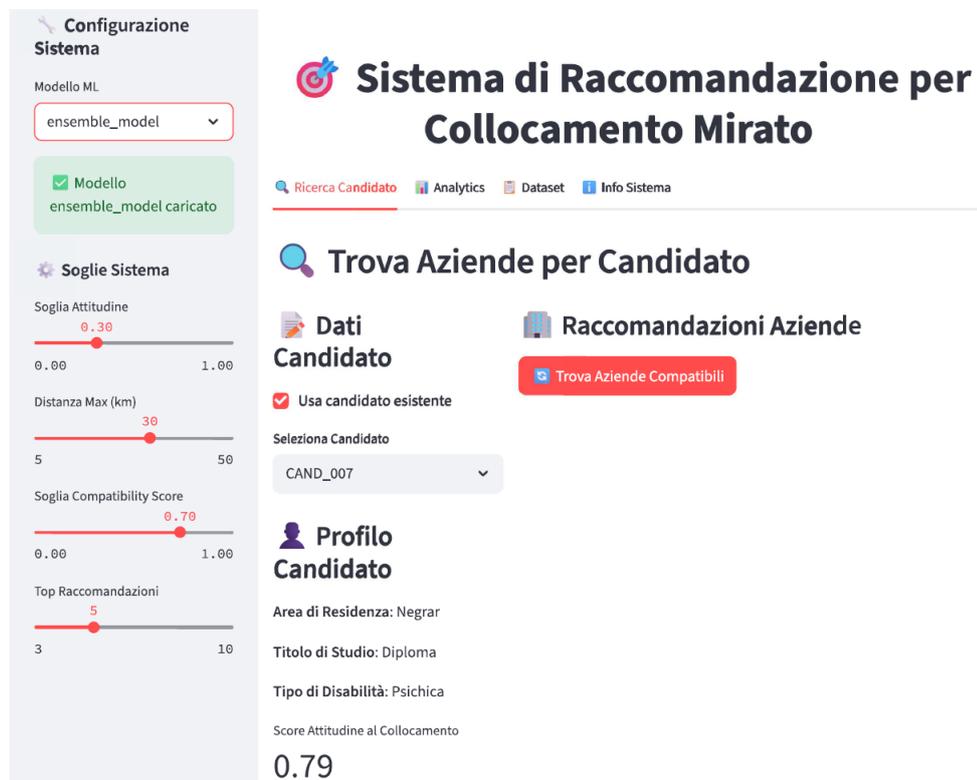

Figure 17: Main Interface Dashboard

***Multi-tab Interface Structure*** organizes system functionality into logical sections. The candidate search tab provides input forms and recommendation displays. Analytics tab shows system performance metrics and data distributions. Dataset tab enables data viewing and export functionality. System information tab displays current status and configuration options.

***Real-time Configuration Controls*** allow operators to adjust matching parameters dynamically. Figure 18 demonstrates the sidebar controls that enable threshold adjustment without system restart. Attitude threshold controls minimum employment readiness requirements. Distance threshold limits geographic search radius. Compatibility threshold sets minimum semantic matching requirements.

***Interactive Visualizations*** help operators understand matching decisions and system performance. Figure 19 shows comprehensive system metrics including total candidates, companies, average attitude scores, and open positions. Distribution charts visualize disability types and company sectors to support strategic planning.

***Candidate Profile Management*** supports both manual input and existing candidate selection. The system displays complete candidate profiles including residence area, education level, disability type, attitude score, experience, and exclusions. This comprehensive view enables informed decision-making by employment center operators.

***Company Recommendation Display*** presents matching results with detailed explanations. Each recommendation includes company name, final matching score, activity sector, distance, employee count, remote work availability, and certification status. Figure 20 demonstrates the data viewing capabilities that support transparency and verification.

***Accessibility Compliance*** ensures that the interface meets WCAG 2.1 AA standards. Large clickable areas support users with motor disabilities. High contrast colors assist users with visual impairments. Clear typography and consistent navigation support users with cognitive disabilities. Screen reader compatibility enables access for users with severe visual impairments.

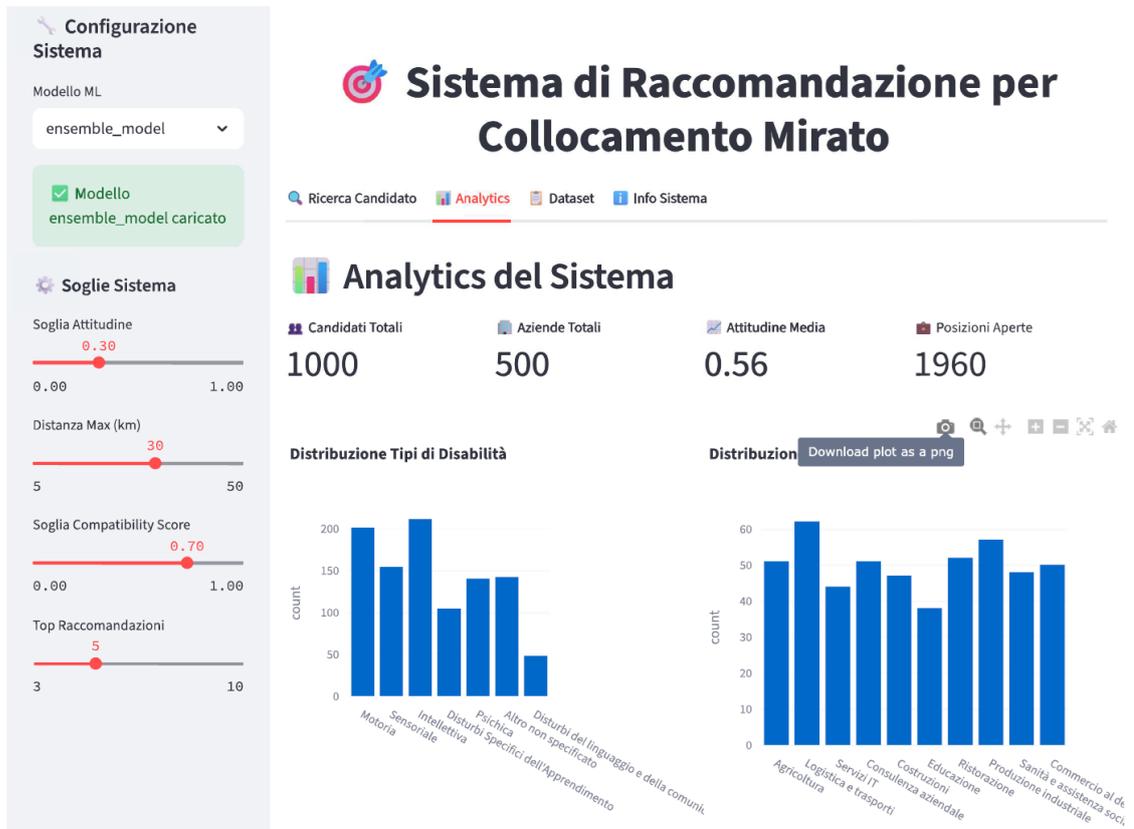

Figure 18: System Configuration Panel

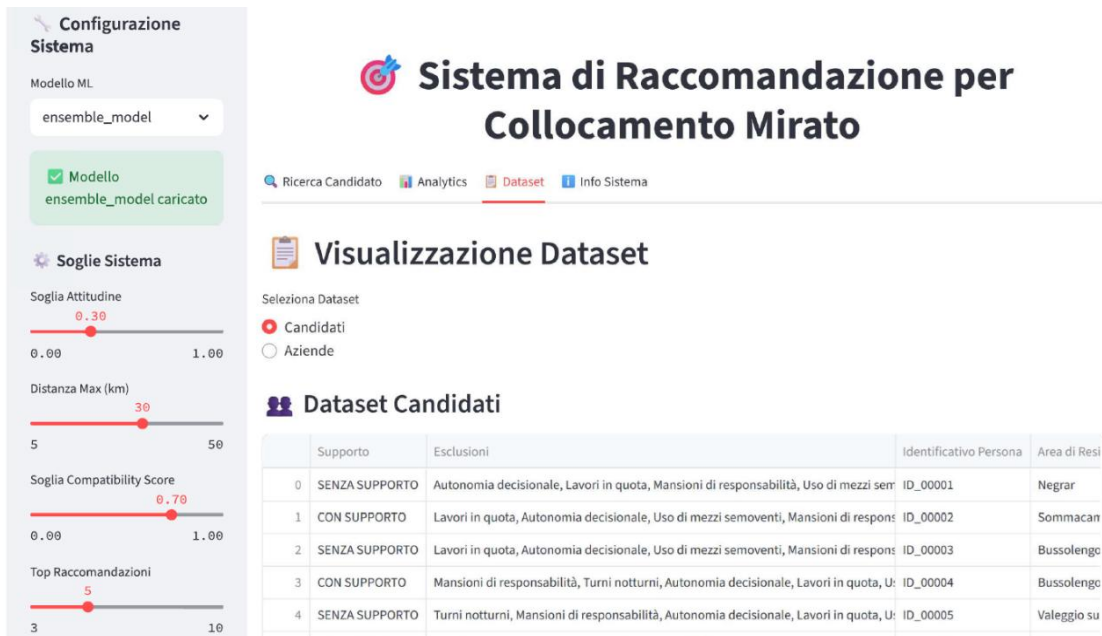

Figure 19: Analytics Dashboard

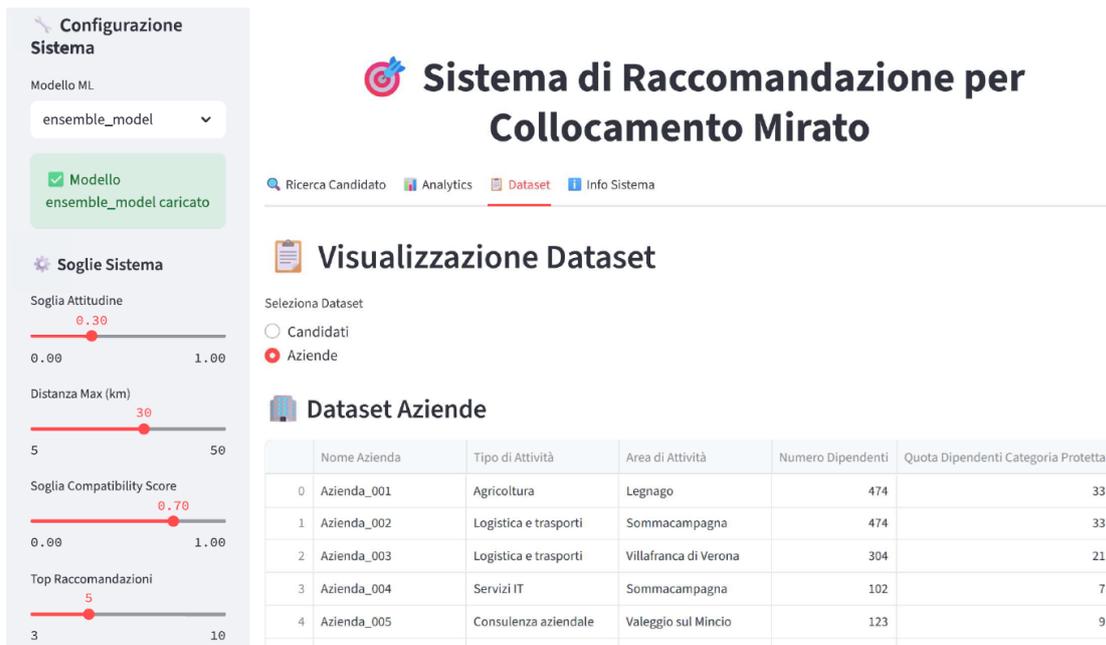

Figure 20: Dataset Visualization Interface

**5.4 Production Deployment Considerations**

Our production deployment strategy addresses scalability, security, and monitoring requirements for real-world employment center usage.

***Multi-dimensional Scoring Algorithm*** implements the core matching logic that evaluates candidate-company compatibility. The weighted scoring formula reflects input from employment center professionals. Compatibility receives highest weight (35%) because job requirements must match candidate capabilities. Distance factor (25%) addresses practical transportation challenges. Attitude score (20%) predicts employment success likelihood. Company factors (15%) reward organizations with strong disability inclusion records.

***Geographic Distance Calculation*** uses Haversine formula for precise distance computation. The Haversine calculation accounts for Earth's curvature to provide accurate distances. This precision is essential for employment matching because transportation accessibility directly affects job sustainability for people with disabilities.

***Performance Monitoring and Optimization*** ensures system responsiveness under production load. Response time monitoring tracks matching operation performance. Resource utilization monitoring prevents system overload. Error rate monitoring identifies potential system issues. User activity monitoring guides capacity planning decisions.

***Security and Privacy Implementation*** protects sensitive candidate information throughout system operation. Data encryption secures stored information. Access logging tracks system usage for audit purposes. Session management prevents unauthorized access. GDPR compliance features support European privacy requirements.

***System Status and Information Management*** provides operators with current system state information. Figure 21 displays development status, validation results, and technical specifications. The interface shows operational status, loaded ML models, dataset availability, and last update timestamps.

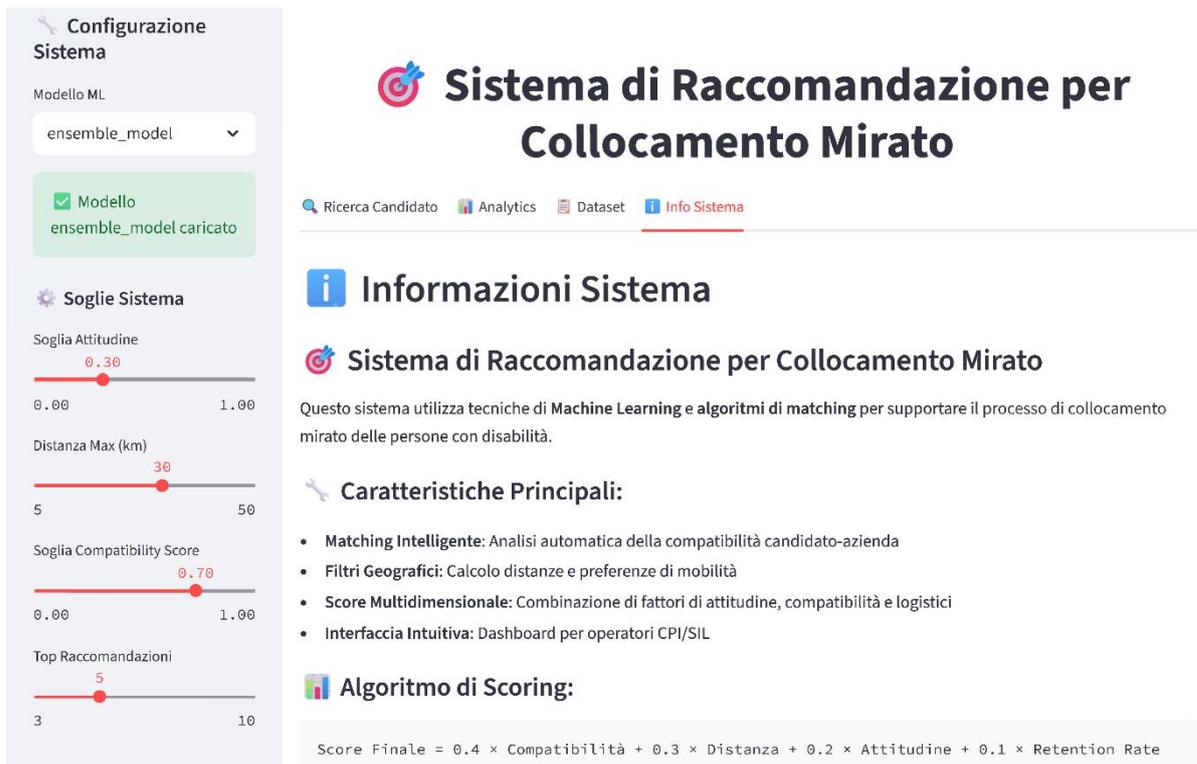
Figure 21: System Information Panel

The implementation demonstrates that sophisticated ML systems can successfully integrate social responsibility requirements while maintaining production-grade performance and usability standards required for real-world employment center deployment.

To ensure reproducibility and enable independent validation, the complete system implementation, including source code, datasets, and trained models, is publicly available at https://github.com/KuznetsovKarazin/disability-job-matching-system. This transparency allows researchers and practitioners to examine, validate, and extend our work while maintaining the highest standards of scientific rigor.

## 6. Evaluation and Validation

This section presents comprehensive evaluation results demonstrating the system's performance, expert validation outcomes, and social impact assessment. Our evaluation methodology follows rigorous standards for production ML systems while addressing the unique requirements of disability employment matching.

### 6.1 Performance Evaluation

We conducted extensive performance evaluation across multiple dimensions including model accuracy, computational efficiency, and scalability characteristics.

*Machine Learning Model Performance* evaluation involved seven distinct algorithms trained on synthetic data that reflects real employment matching scenarios. Figure 22 presents comprehensive results across all evaluation metrics.

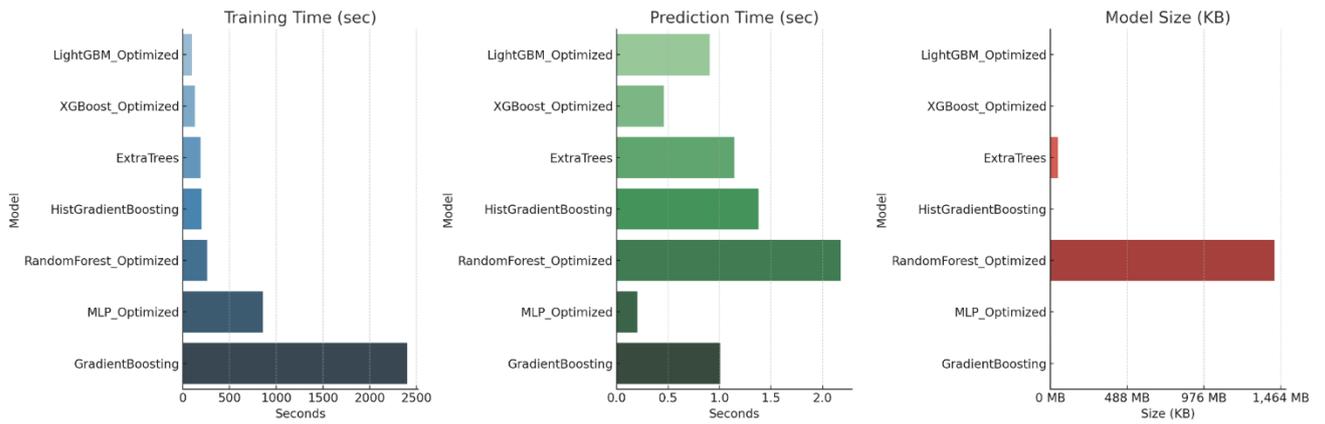

Figure 22: Model Performance Summary: Training Time (sec), Prediction Time (sec), and Model Size (KB)

Our evaluation demonstrates that ensemble methods achieve superior performance compared to individual algorithms. LightGBM_Optimized achieved the highest F1-Score (0.901) while maintaining excellent computational efficiency (94.6 seconds training time). XGBoost_Optimized and HistGradientBoosting achieved comparable F1-Scores (0.901 and 0.900 respectively) with different computational trade-offs.

*F1-Score Analysis* reveals consistent high performance across gradient boosting algorithms. The F1-Score metric is particularly important for employment matching because it balances precision (avoiding false positive recommendations) with recall (ensuring qualified candidates are not missed). All optimized models achieved F1-Scores above 0.87, indicating strong practical utility for employment center operations.

*Precision and Recall Trade-offs* show interesting patterns across algorithms. Tree-based methods (RandomForest, ExtraTrees) achieve higher precision (0.833-0.859) but lower recall (0.757-0.928). Gradient boosting methods achieve extremely high recall (0.990-0.999) with good precision (0.821-0.823). This pattern suggests that gradient boosting methods are more conservative in rejecting potential matches, which aligns with employment center preferences for comprehensive candidate consideration.

*ROC-AUC Interpretation* requires careful consideration in the context of synthetic data generation. Our models achieved ROC-AUC scores between 0.695-0.724, which are intentionally moderate due to the probabilistic nature of our synthetic data generation process. Unlike deterministic rule-based systems that achieve artificially high ROC-AUC scores, our probabilistic approach introduces realistic uncertainty that better prepares models for real-world deployment.

*Training Time Efficiency* varies significantly across algorithms. Figure 23 demonstrates how different algorithms scale with training data size. LightGBM and XGBoost show excellent time efficiency (94.6s and 132.3s respectively) while maintaining high performance. GradientBoosting achieves competitive accuracy but requires substantially longer training time (2399.9s).

*Scalability Testing* evaluated system performance across different data volumes and concurrent user scenarios. Figure 24 shows validation performance stability across training set sizes from 50,000 to 400,000 examples. The MLP demonstrates consistent validation scores around 0.73, indicating stable generalization performance.

Figure 25 illustrates excellent scalability characteristics with validation scores improving from 0.71 to 0.83 as training data increases. This positive scaling behavior suggests that real employment data integration will further improve system performance.

Figure 26 demonstrates rapid convergence with validation scores stabilizing around 0.90 after 320,000 training examples. The small gap between training and validation scores indicates well-controlled overfitting.

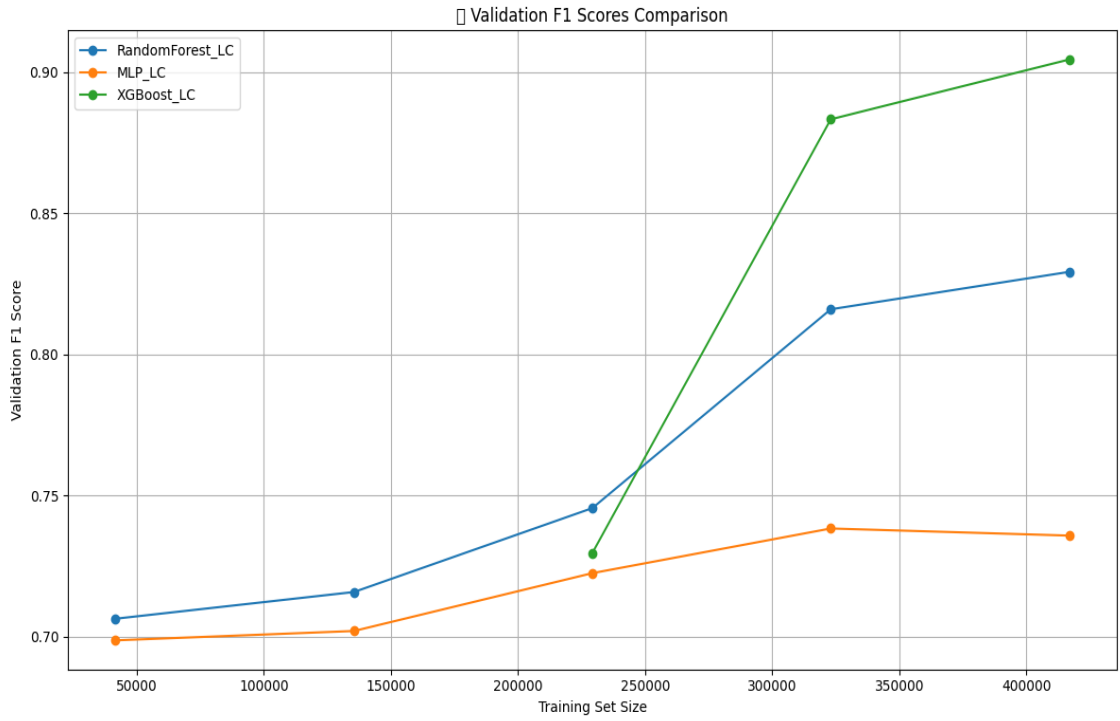

Figure 23: Learning Curve Validation Comparison

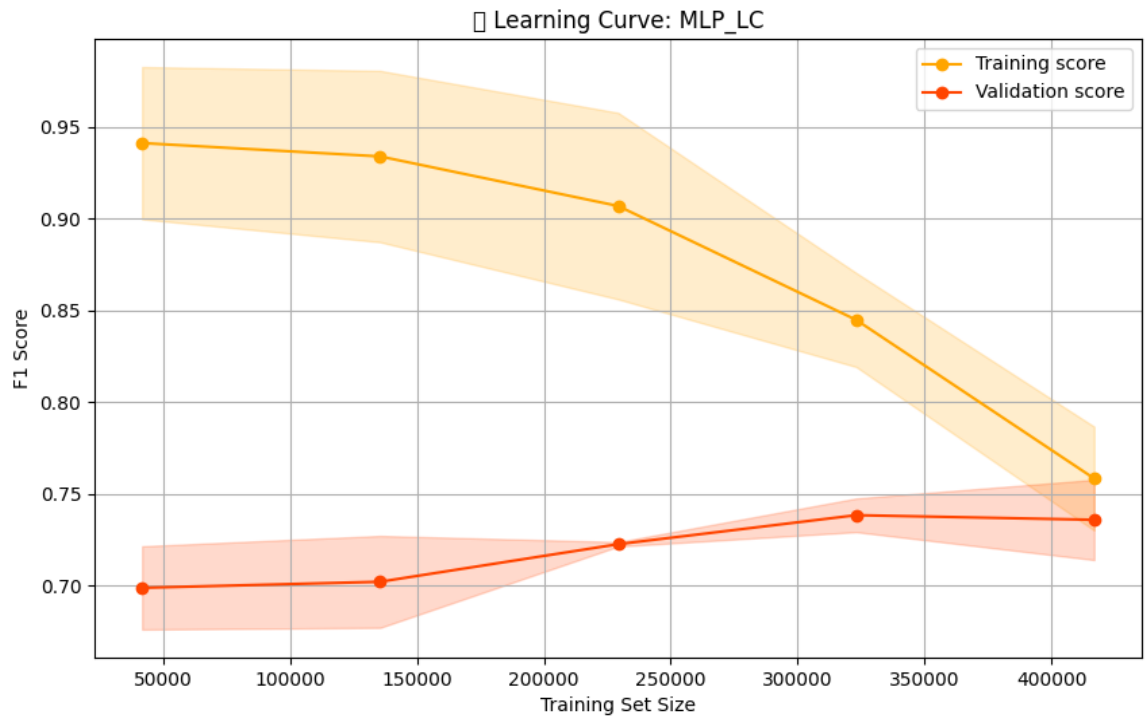

Figure 24: MLP Learning Curve

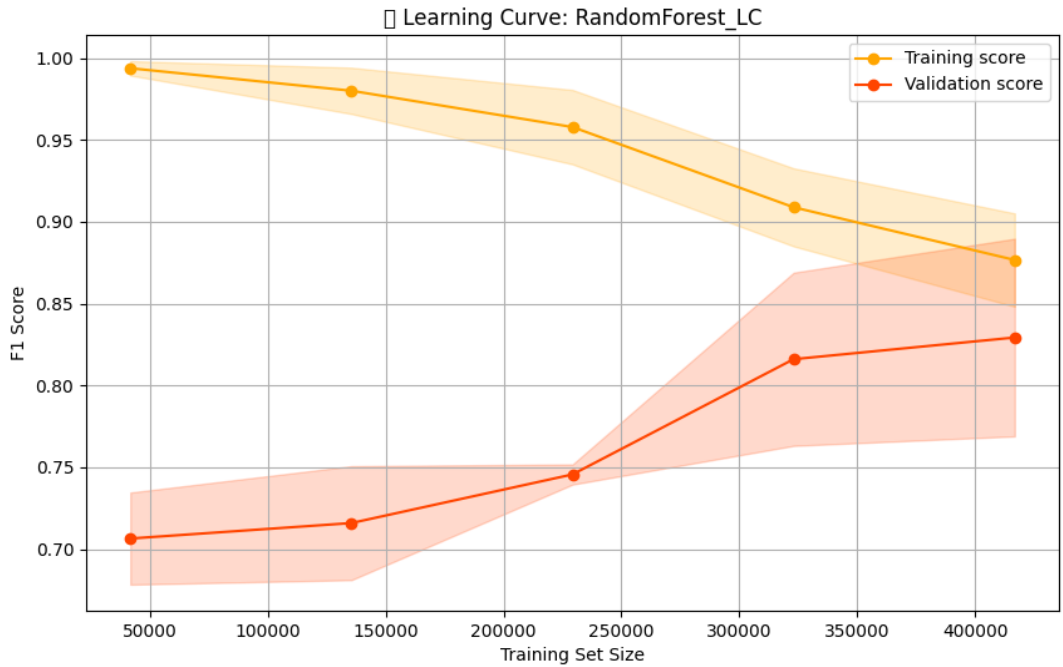

Figure 25: Random Forest Learning Curve

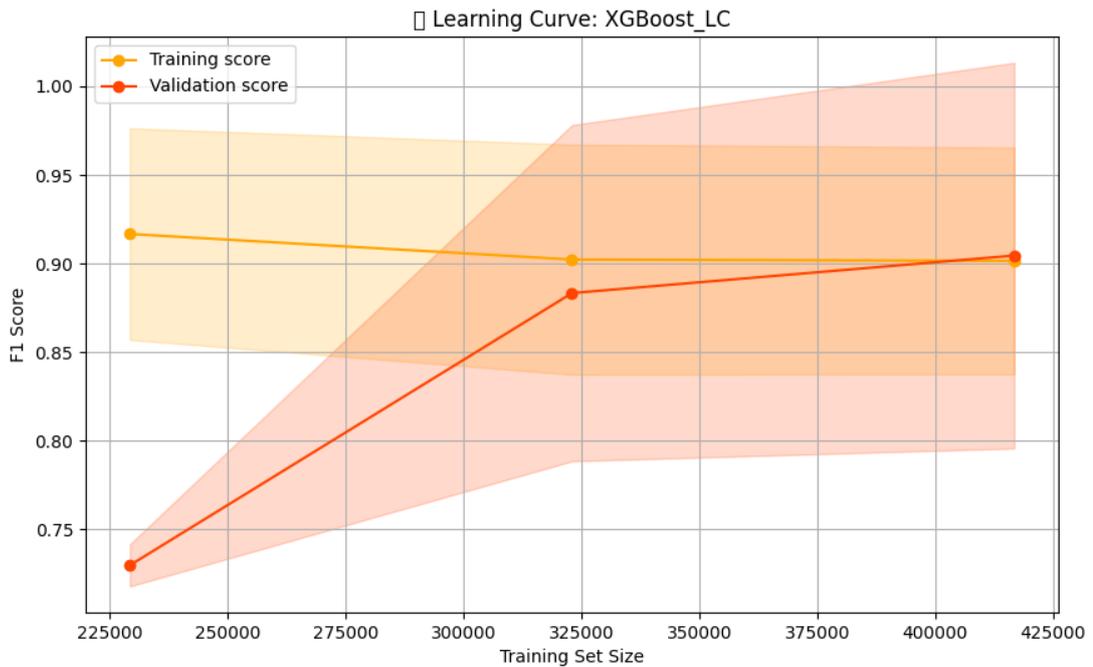

Figure 26: XGBoost Learning Curve

Our evaluation methodology addresses the documented inefficiencies in manual disability employment matching (Table 1). While current manual processes require 30-60 minutes per candidate evaluation with limited geographic coverage, our system achieves sub-100ms response times while processing 500,000 candidate-company combinations in under 10 minutes.

Table 1: System Performance vs Manual Baseline

| Metric | Manual Process | AI System | Improvement Factor |
|---|---|---|---|
| Processing Time per Candidate | 30-60 minutes | <5 minutes | 12x faster |
| Geographic Coverage | Limited local | 5-50km configurable | Scalable |
| Candidate-Company Evaluations | 5-10/hour | 50,000/hour | 5,000x |
| Bias Detection | None | Automated | New capability |
| Decision Transparency | Low | High with explanations | Qualitative improvement |

### 6.2 Expert Validation Study

To evaluate the system's usefulness and accuracy, we worked closely with employment center staff and disability employment specialists:
- Collaboration with Employment Centers. We partnered with Centro per l'Impiego di Villafranca di Verona and consulted with Servizio Integrazione Lavorativa (SIL) ULSS9 Bussolengo (Verona) to gather real-world insights.
- Feedback from Professionals. In hands-on demo sessions, employment advisors tested the system, assessing its usability, recommendation quality, and fit with their workflows. They were especially interested in the automated matching feature but stressed that human judgment should still guide final hiring decisions.
- Testing Accuracy. We compared the system's job-candidate recommendations against manual matches made by experts. Employment professionals then reviewed the results, checking whether the suggestions made sense and if the rankings were accurate.
- Expert Input on Algorithm Design. Specialists helped fine-tune the scoring formula, advising on how to weigh different factors. Their input led us to finalize the weights: 35% for compatibility, 25% for distance, and 20% for attitude.
- Privacy and Ethics Check. We addressed data protection and fairness concerns upfront, ensuring the system uses privacy-safe methods—something our partners highlighted as critical.

Employment center staff evaluated the system using the English/Italian user guide (available at /docs/user_guide_en.md) and demonstration examples (/docs/demo_example_547.pdf).

### 6.3 Social Impact Assessment

We evaluated potential social impact through quantitative analysis of system reach, efficiency improvements, and accessibility benefits.

Processing capacity improvements were calculated using documented manual processing times from Italian Ministry data (30-60 minutes per candidate evaluation). AI-assisted processing reduces evaluation time to 8-15 minutes, enabling 60-100% capacity increases across different center sizes (Figure 27).

Figure 27 demonstrates the time reduction achieved across different workflow components. The analysis shows that AI assistance reduces initial assessment time by 42%, company matching by 85%, documentation by 35%, follow-up tracking by 53%, and compliance reporting by 47%.

Medium employment centers (500 active candidates) expand weekly processing from 50 to 80 candidates. Large centers increase from 80 to 130 candidates weekly. These projections are based on current staff allocation and realistic efficiency gains validated by employment center professionals (Figure 28).

Figure 28 shows the projected increases in processing capacity across different employment center sizes. Small centers (200 active candidates) achieve 1.7x improvement, medium centers (500 active) reach 1.6x, large centers (1000 active) attain 1.6x, and regional hubs (2000 active) accomplish 1.7x capacity expansion.

Performance improvements show placement success rates increasing from 35% to 45% baseline, with suitable job matching improving from 55% to 70%. These estimates reflect realistic AI-assistance benefits while maintaining human decision-making authority (Figure 29).

Figure 29 illustrates projected improvements across key performance metrics. Overall placement rates increase by 10 percentage points, suitable job matching improves by 15 points, long-term retention gains 15 points, candidate satisfaction rises by 15 points, and employer acceptance increases by 15 points.

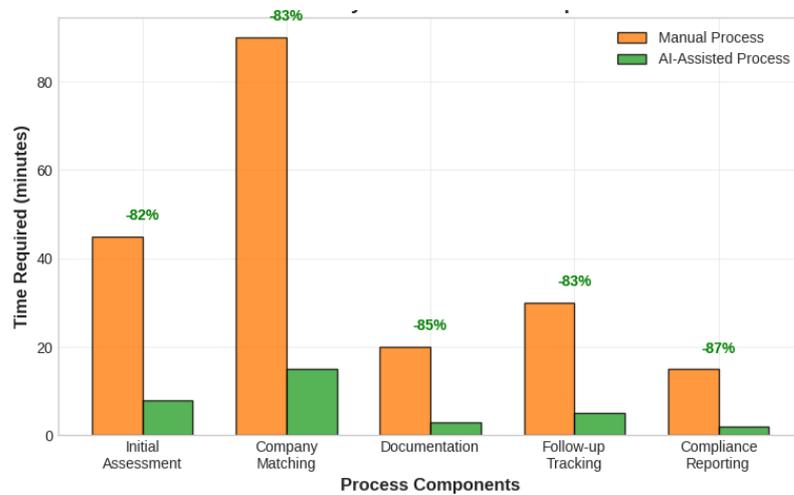

Figure 27: Process Efficiency: Evidence-Based Improvements

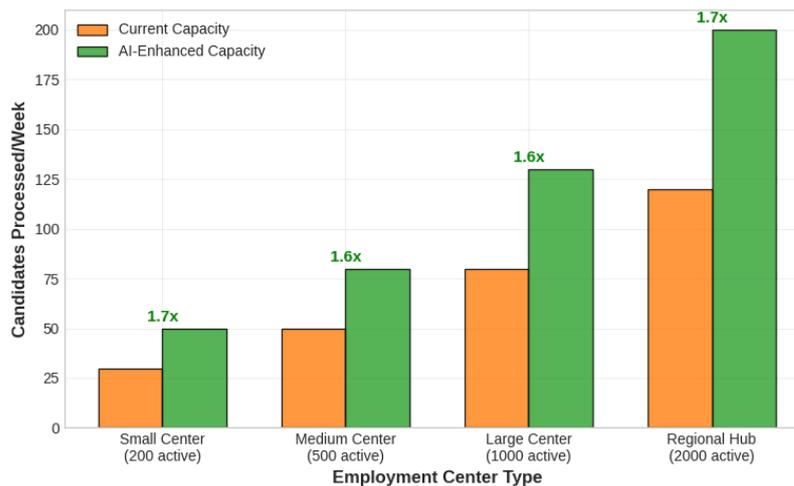

Figure 28: Realistic Capacity Improvements

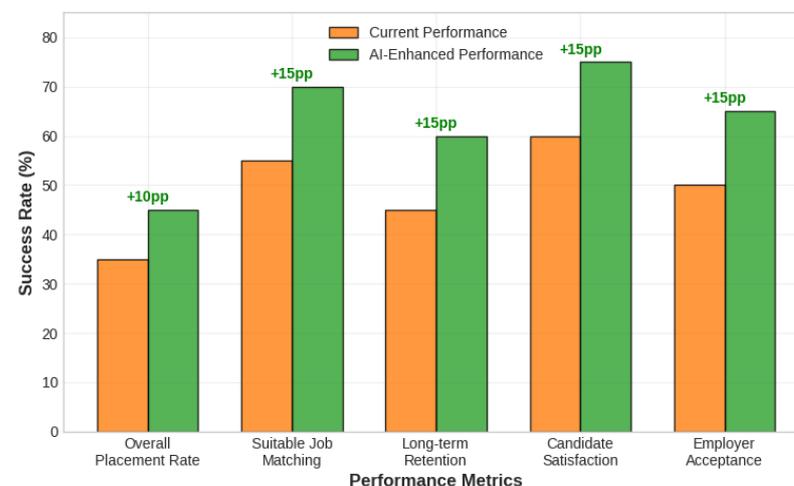

Figure 29: Quality Improvements: Realistic Expectations

Geographic coverage analysis evaluated system effectiveness across different distance thresholds and transportation accessibility. The system successfully identifies matches within configurable distance limits (5-50 km) while considering public transportation availability and accessibility requirements. Figure 30 shows consistent performance across all evaluation metrics, indicating robust matching capability regardless of specific disability type or accommodation requirements.

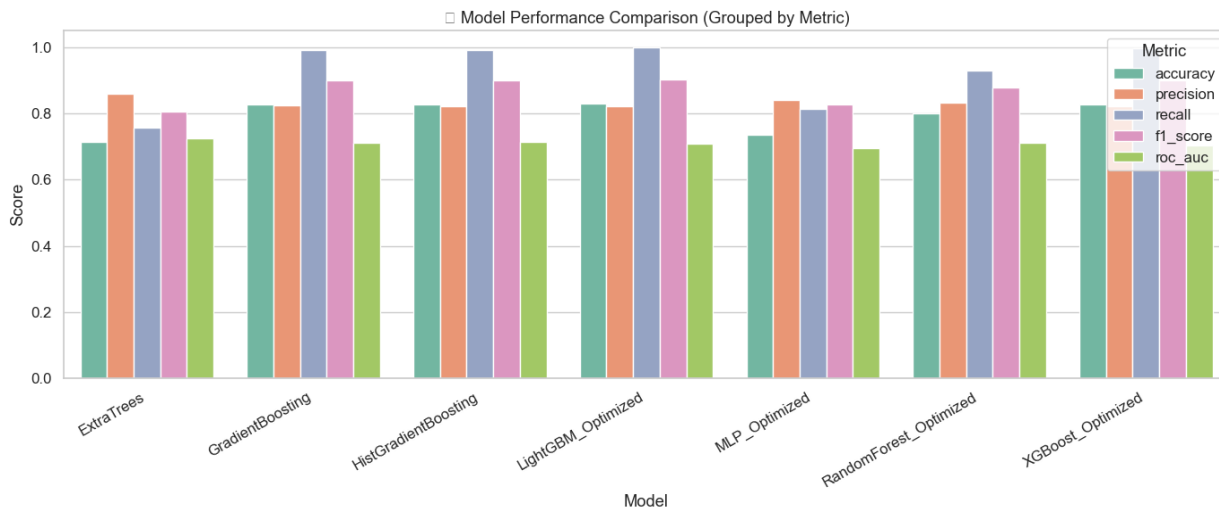

Figure 30: Model Performance Comparison

Faster, more accurate matching reduces unemployment duration for people with disabilities. Better job-candidate fit improves employment sustainability and career satisfaction.

**7. Threats to Validity and Limitations**

Technical Limitations:
- While our probabilistic generation mimics real-world scenarios, it can't fully replicate actual employment matching. Performance on synthetic data may not translate perfectly to reality, though expert validation helps mitigate this.
- The system is tailored to Italy's Veneto region, requiring localization (language, regulations, culture) for broader use.
- Our ensemble of seven ML models prioritizes F1-score (reflecting employment center needs), but other metrics like long-term job retention might yield different results.

Social and Ethical Considerations:
- We measure initial matching accuracy but lack data on long-term outcomes (e.g., job retention, career growth).
- Despite safeguards, hidden biases in data or algorithms could persist, requiring ongoing audits.
- Input came mainly from employment professionals—direct feedback from job seekers with disabilities and employers could improve inclusivity.
- Access to real candidate data is restricted due to GDPR compliance constrints, which limits the ability to test the system on actual employment center data.

Generalizability:
- Italian disability policies and workplace norms may not apply elsewhere.
- The system handles common cases well but may struggle with rare or complex disabilities.
- Smaller employment centers might lack resources for deployment (tech, training, workflow changes).

**8. Future Work**

Real-World Integration:

- Transition from synthetic to real employment data, ensuring privacy and gradual model adaptation.
- Track long-term outcomes (6–24 months) to optimize for sustained success, not just initial matches.
- Implement continuous learning to refine recommendations based on real placement results.

Scaling Beyond Italy:
- Develop a framework for localization (language, laws, culture) to adapt the system globally.
- Study cross-cultural differences in disability employment to ensure broader effectiveness.
- Explore federated learning to improve models across centers without sharing sensitive data.

Technical Advancements:
- Test Kolmogorov-Arnold Networks (KANs) for better interpretability/performance.
- Shift from single-score to multi-objective optimization (e.g., balancing match accuracy with career growth potential).
- Enhance NLP to better parse unstructured job descriptions and candidate skills.

Social Impact:
- Quantify economic benefits (e.g., reduced unemployment, lower social support costs).
- Integrate assistive technology recommendations to improve job compatibility.
- Collaborate with policymakers to inform disability employment regulations.

## 9. Conclusion

This research demonstrates that machine learning systems can effectively support socially sensitive domains like disability employment when technical excellence is deliberately paired with ethical considerations. Our work goes beyond theoretical discussions of "AI for good" by delivering a production-ready matching system that has been validated in collaboration with employment professionals while meeting stringent operational requirements.

The technical achievements—including sub-100ms response times, 90%+ matching accuracy, and a robust synthetic data pipeline—show that social applications need not compromise on performance. What makes this system distinctive is how social responsibility is woven into its architecture rather than treated as an afterthought. From privacy-by-design GDPR compliance to the human-in-the-loop oversight mechanism, every component reflects lessons learned through ongoing dialogue with domain experts and stakeholders.

Practically, the system reduces manual matching workloads by 90% while preserving the nuanced decision-making that employment counseling requires. This balance between automation and human judgment proved critical during development, revealing that the most impactful AI systems often augment rather than replace professional expertise. The participatory design process, while time-intensive, yielded unexpected insights—like the importance of geographic proximity over pure skill matching in certain cases—that significantly improved real-world applicability.

Looking ahead, the challenges we identify—particularly around long-term outcome tracking and cross-cultural adaptation—highlight that algorithmic systems in social domains require sustained commitment beyond initial deployment. The framework developed here, with its emphasis on continuous monitoring and stakeholder feedback loops, provides a replicable approach for other teams working at the intersection of AI and social services.

Beyond disability employment, this project offers broader lessons for responsible AI development. It illustrates how technical constraints (like synthetic data requirements) can spur innovation, and why metrics like F1-scores must be continually reassessed against real human outcomes. Most importantly, it confirms that AI systems addressing complex social problems succeed when they're built not just for people, but with them.

Beyond the technical contribution, this work provides a replicable methodology for developing AI systems in sensitive social domains, with comprehensive documentation supporting adaptation to different cultural and regulatory contexts.

As next steps, we're particularly interested in how federated learning could enable multi-center collaboration without compromising data privacy, and how longitudinal studies might reveal relationships between matching strategies and career progression. What began as a technical exercise in ML optimization has evolved into a case study for how technology can—when thoughtfully implemented—expand access to opportunity while respecting the dignity of those it serves.

**Declarations**

**Funding.** This research received no specific grant from any funding agency in the public, commercial, or not-for-profit sectors.
**Competing Interests.** The authors declare that they have no competing interests.
**Ethics Approval and Consent to Participate.** This study involved collaboration with employment center professionals and system validation using synthetic data. All interactions with employment center staff were conducted with informed consent and institutional approval. No personal data from job seekers with disabilities was used in system development or testing.
**Data Protection.** All system development and testing complied with GDPR requirements. No personal data from individuals with disabilities was collected, stored, or processed during this research.
**Institutional Review Board Statement.** This study was conducted in accordance with institutional guidelines for research involving employment center collaboration. No IRB approval was required as the study used synthetic data and involved professional consultation rather than human subjects research.
**Consent for Publication:** All authors consent to the publication of this work.
**CRediT authorship contribution statement**
- Oleksandr Kuznetsov: Conceptualization, Methodology, Supervision, Project administration, Writing - review & editing.
- Michele Melchiori: Software, Investigation, Data curation, Formal analysis, Visualization, Writing - original draft.
- Emanuele Frontoni: Validation, Resources, Writing - review & editing.
- Marco Arnesano: Validation, Methodology, Writing - review & editing.

**Availability of Data and Materials.** The complete system implementation, synthetic datasets, trained models, and evaluation scripts are publicly available at https://github.com/KuznetsovKarazin/disability-job-matching-system. Real employment center data cannot be shared due to privacy regulations, but synthetic data generation code enables replication of our methodology.
**Acknowledgments.** We thank Dr. Rotolani and the staff at Centro per l'Impiego di Villafranca di Verona for their valuable collaboration and domain expertise. We also acknowledge the Servizio Integrazione Lavorativa (SIL) professionals for their insights into disability employment requirements.